\documentclass[]{IEEEtran}

\usepackage{amsmath}
\usepackage{amsfonts}
\usepackage{amssymb}
\usepackage{makeidx}
\usepackage{cite}
\usepackage{graphicx}
\usepackage{mathtools}
\usepackage{cases}
\usepackage{xcolor}
\usepackage{multicol}
\usepackage{cite}

\usepackage{hyperref}
\usepackage{bbm}
\usepackage[normalem]{ulem}
\usepackage{bigints}
\usepackage{braket}

\newtheorem{theorem}{Theorem}
\newtheorem{corollary}{Corollary}

\newtheorem{example}{Example}

\newtheorem{remark}{Remark}

\DeclareMathOperator*{\argmax}{arg\,max}

\DeclareMathOperator{\cW}{\mathcal{W}}

\DeclareMathOperator{\cD}{\mathcal{D}}

\DeclareMathOperator{\cL}{\mathcal{L}}

\DeclareMathOperator{\cT}{\mathcal{T}}
\DeclareMathOperator{\cR}{\mathcal{R}}

\DeclareMathOperator{\cK}{\mathcal{K}}

\DeclareMathOperator{\SIR}{\textrm{SIR}}

\DeclareMathOperator{\bP}{\mathbf{P}}
\DeclareMathOperator{\ind}{\mathbbm{1}}
\DeclareMathOperator{\bE}{\mathbf{E}}
\DeclareMathOperator{\bZ}{\mathbb{Z}}
\newcommand*\diff{\mathop{}\!\mathrm{d}}

\newcommand*\nnb{\nonumber}
\newcommand*\nnnl{\nonumber\\}

\DeclarePairedDelimiter{\floor}{\lfloor}{\rfloor}

\newcommand{\uta}[1]{\underbrace{#1}_{\text{(a)}}}
\newcommand{\utb}[1]{\underbrace{#1}_{\text{(b)}}}

\newcommand{\ea}{\stackrel{(\text{a})}{=}}
\newcommand{\eb}{\stackrel{(\text{b})}{=}}
\newcommand{\ec}{\stackrel{(\text{c})}{=}}
\newcommand{\ed}{\stackrel{(\text{d})}{=}}

\title{Modeling and Analysis of Data Harvesting Architecture based on Unmanned Aerial Vehicles}

\author{Chang-Sik~Choi, Fran{\c{c}}ois Baccelli, and Gustavo de Veciana
\thanks{Chang-sik Choi is a senior system engineer of Qualcomm Inc, Bridgewater, New Jersey. Francois Baccelli is with Department of Electrical and Computer Engineering and Department of Mathematics, The University of Texas at Austin, TX, USA. Gustavo de Veciana is with Department of Electrical and Computer Engineering, The University of Texas at Austin, TX, USA. (emails: chang-sik.choi@utexas.edu, baccelli@math.utexas.edu, gustavo@ece.utexas.edu)}
\thanks{A short version is in Proc. IEEE ISIT 2019 \cite{choi2019ISIT} }}
\begin{document}
	\maketitle
\begin{abstract}
This paper explores an emerging wireless Internet-of-things (IoT) architecture based
on unmanned aerial vehicles (UAVs). We consider a network where a fleet of UAVs at a
fixed altitude flies on planned trajectories and IoT devices on the ground are scheduled
to transmit their data to the UAVs when the latter are nearby. In such a system, the UAVs'
motion triggers the uplink transmissions of the IoT devices. As a result, network performance
is determined by the geometric and dynamic characteristics of the system. We propose a joint
stationary model for UAVs and IoT devices and then evaluate the interference, 
the coverage probability, and the data rate of the typical UAV. To assess the harvesting
capability of the proposed architecture, we derive a formula for the amount of data uploaded
from each IoT device to a UAV. We also establish a linear relationship between the UAV coverage
and the harvesting capability of the network, which provides insights into the design of the
proposed harvesting scheme. In addition, we use our analytical results to numerically 
show that there exists a trade-off between the uploaded data and the size of the IoT
scheduling window. Specifically, for a given UAV and IoT geometry, there exists an
optimal scheduling window that maximizes the harvesting capability of the proposed network. 
\end{abstract}
\section{Introduction}\label{S:1}
\subsection{Motivation and Related Work}
This paper investigates a wireless data harvesting architecture for static ground-based
Internet-of-Things devices (IoT) \cite{gubbi2013internet,da2014internet,7582463}. 
IoT devices may have various functions such as performing environmental measurements
from some area, measuring electric and water systems, collecting average road traffic counts,
and monitoring extreme environments\cite{baiocchi2013uav,gaur2015smart,botta2016integration}.
The data continuously generated from these IoT devices are uploaded to the Internet
using technologies such as narrowband-IoT \cite{ratasuk2016nb,wang2017primer}
or low-power-wide-area networks\cite{sinha2017survey,raza2017low,guo2017massive}.
Currently, these approaches rely on the existence of fixed network infrastructure, e.g.,
base stations or access points. Consequently, the performance of such technologies
is often limited by the bottlenecks around infrastructure nodes \cite{1402592,Choi:2018:DLM:3209582.3209590}.
When the density of IoT devices varies across areas, deploying more base stations may be
required for improving performance. Installing denser infrastructure might be impractical
due to technological and financial reasons.

One possible alternative to collect IoT data without increasing the fixed network infrastructure
is to leverage mobile agents, e.g., vehicles 
\cite{jain2006exploiting,Xing:2008:RDA:1374618.1374650,forstall2013mobile,saad2014vehicle,talluri2018enhanced,Choi:2018:DLM:3209582.3209590,8796442} 
which visit IoT devices and collect their data. When the vehicles arrive at stations connected to the network,
the data will be uploaded to the Internet. This vehicle harvesting concept requires a
delay-tolerant setting, where some delay is accepted
\cite{grossglauser2002mobility,zhao2005controlling,balasubramanian2007dtn,fall2008dtn,zorzi2003geographic,skordylis2008delay,abdrabou2011probabilistic} in order to improve network capacity or reduce aggregate interference.
As discussed in \cite{gubbi2013internet,da2014internet,7582463}, a number of 
IoT applications tolerate such delays. For such applications,
relying on mobile vehicle harvesters is a viable and economical option to collect IoT data.

Nevertheless, IoT data harvesting based on mobile vehicles suffers of a few practical restrictions.
For instance, vehicles travel only on the existing roads and thus some IoT devices---such as 
those with short transmission range or located in rural areas with sparse road networks---may not
be able to establish reliable connections  to nearby vehicles at any time
\cite{choi2018Globecom,choi2018poisson}. As an alternative to vehicle harvesters, 
UAVs are considered as a way to extend mobile data harvesting beyond the existing
road systems \cite{chandrasekharan2016designing,lin2018sky,muruganathan2018overview}.
In practice, certain regulations restrict the use of UAVs. For instance, UAVs operated for recreational
purposes are not allowed to enter no-flying zones or to operate near groups of people,
public events, or a stadium. In addition, in the United States, pilots of UAV operated for commercial usage
must be certified by the Federal Aviation Administration \cite{cracknell2017uavs}.
Even when these restrictions apply, UAV configured to meet the needs of telecommunication operators
could become an effective way to achieve large-scale connectivity with a marginal infrastructure
support, particularly so in areas with a sparse road system and limited wired backhaul networks.
Here are a few concrete examples of such UAV networks.
\begin{itemize}
	\item A fleet of UAVs monitoring a large area where small sensor devices are constantly collecting seismic activity data \cite{baiocchi2013uav};
	\item A platoon of military UAVs flying over a wide area to collect scattered information from IoT sensors on the battle ground \cite{6761569};
		\item Intelligent transportation system UAVs flying on certain streets and  harvesting traffic or sensor data from infrastructure such as roadside units \cite{7555867,botta2016integration}.
\end{itemize}
In these examples and many others, the network elements consist of UAVs and static data devices.
These data devices are deployed over a large area where fixed infrastructure is unavailable or undesirable. 
These sensors or devices persistently produce data and a UAV fleet is operated to collect them.

The literature on wireless networks relying on UAVs is quite rich, with contributions
stemming both from industry stakeholders and academic researchers
\cite{6761569,chandrasekharan2016designing,muruganathan2018overview,8254658,8053918,lin2018sky,hattab2018energy,zeng2019accessing,7555867,8709739,8488493}.
Stochastic geometry was used to study the performance of UAV-assisted cellular networks
and device-to-device networks in \cite{7967745,7994915,7412759,8377426}.
UAV downlink/uplink performance was also analyzed in \cite{8254658,8377426}.

Compared to this literature, the first novelty of the present paper 
is the new parametric stochastic geometry model allowing one to represent architectures based on fleets
of operated UAVs. This is probably the simplest, though not simplistic, parametric model in this class
as UAV motion is linear and with constant velocity rather than adaptive and linked to traffic.
The second novelty is the analytic framework based on the evaluation of the interference field statistics
which allows us to derive expressions for various classical metrics, such as the probability of coverage,
and also for less classical ones such as the mean amount of data harvested from each IoT device
as UAVs progress. To the best of our knowledge, the last metric was never been analyzed.
The third novelty is the optimization of the harvesting capability as a function of the network
parameters.

\subsection{Contributions}

\textbf{A parametric analytical framework for UAV harvesting networks}: We propose a parametric
model for data harvesting architectures where UAVs collect delay-tolerant data from
static surface-level IoT devices. UAVs are designed to cover a wide area, with no need of
fixed infrastructure, e.g., base stations. At any time, every UAV receives data
transmission from one of the IoT devices located in a rectangle activation window
of size $ w $ by $ l $ on the ground. The fleet of UAVs is assumed to move at constant 
speed $ v $ in a coordinated way. As the fleet of UAVs progresses, the windows of
UAVs also progress on the ground. The parameters as set to values such that the
proposed architecture provides universal coverage of IoT devices, that is, each device
is periodically covered by some UAV (rather than always covered as in the fixed
infrastructure architecture case). Parametric stationary point processes are used to
model the locations of IoT devices and UAVs, respectively. Specifically, the locations
of the UAVs are modeled by a stationary grid characterized by inter-point distances $\mu$
between UAVs on any given line and some common altitude $h$.
The locations of IoT devices on the ground are modeled
by an independent planar Poisson point process with a fixed intensity.

\textbf{Performance analysis}: We first use Palm calculus to derive the Laplace transform
of the interference seen by a typical UAV. We show that interference distribution is time-invariant.
We also derive the distribution of the signal-to-interference (SIR) ratio of the typical UAV and 
the distribution of its data rate. Similarly, we obtain an expression for the mean value
of the amount of harvested data transmitted from the typical IoT device to a UAV
when the latter passes over the IoT device location. We use the mass transport principle to link
the mean value of the harvested data and the instantaneous UAV data rate.

\textbf{Network performance trade-off} As explained above the fact that UAVs are in motion,
allows such an architecture to provide universal (though intermittent) coverage 
with a smaller number of UAVs than what would be needed in a static architecture based on infrastructure nodes.
(specifically, in the system model, the window size $ w $ is set to be less than $ \mu $).
Furthermore, thanks to inherent mobility, UAVs are close to the IoT devices when transmissions occur,
increasing the received signal power. Similarly, UAVs can be operated at a safe distance from each other,
reducing the interference. In this sense, the window size $ w $ is one of the key parameters that
characterize the performance of the proposed motion-based harvesting architecture. We show that
there is a trade-off relationship between the data rate of the typical UAV and the size of the window.
More precisely, for all given densities of IoT devices, there exists a unique value $ w^\star $
that maximizes both the data rate of the typical UAV and the harvesting capability of the network.

\section{System Model}\label{S:2}

This section introduces the system model of the proposed network architecture.
Then, access control, propagation model, and network performance metrics are discussed.
In order to ease the exposition, we start with the linear case (a single line
of UAVs progressing a speed $v$ along this line, and a strip of IoT devices along
this line) and postpone the description of the planar case (an infinite number of parallel 
lines covering the plane, each with UAVs as above, with a homogeneous planar point process of
IoT devices) to the end of the paper.

\begin{figure}
	\centering
	\includegraphics[width=1\linewidth]{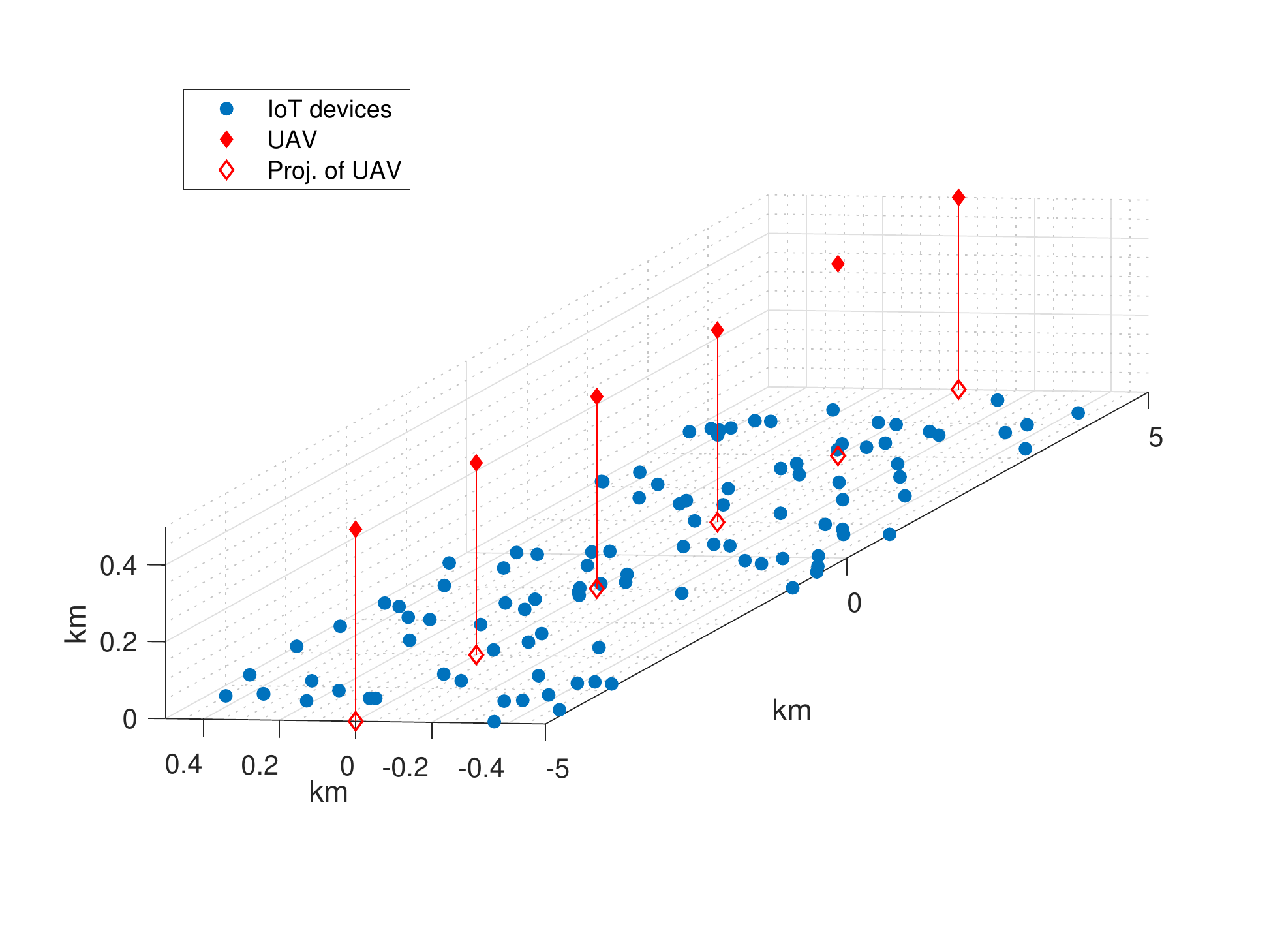}
	\caption{Illustration of the proposed IoT harvesting network where the UAVs are located at the altitude $ z=0.5 $ km and the distance between UAVs is given by $ 2 $km. The platoon of UAVs is assumed to move in the positive $ x$-axis with speed $ v. $ The activation window of each UAV is characterized by a rectangle $ w\times l $ centered at the projection of the UAV on the plane. See Fig. \ref{fig:tdmapolicy} for the activation window.}
	\label{fig:dronesandiots}
\end{figure}
\subsection{Moving UAV Fleet and Stationary IoT Devices}\label{S:2A}
As mentioned in the introduction, this paper focuses on the data-harvesting applications of a UAV fleet with a idealized motion planned by some operator. These UAVs are separated by some distance while in the air to avoid collision. Thus, we use periodic points on a line parallel to the $ x $-axis in order to model the UAVs in the air. The distance between the points is given by $ \mu $ and the altitude of the UAVs is given by $ h. $ To address the randomness of the UAVs at time $ 0 $, the periodic points are shifted by a single uniform vector, $ (U,0,0) $ where $ U \sim \text{Uniform}\left[-\frac{\mu}{2},\frac{\mu}{2}\right]$. 
\begin{equation}\label{10}
\Psi(0)=\sum_{k\in\bZ}\delta_{(k\mu,0,h)+(U,0,0)},
\end{equation} 
where $ \delta_x $ denotes the Dirac measure indicating a point mass at location $ x$, and $ \bZ $ denotes the set of all integers. This model is probably the simplest model for the planned movements of a UAV fleet. For analytical tractability, we assume that all UAVs move at the same speed $ v $ in the positive $ x$-axis. Then, the locations of UAVs at time $ t>0 $ are given by 
\begin{equation}\label{11}
\Psi(t)=\sum_{k\in\bZ}\delta_{(k\mu,0,h)+(U,0,0)+(vt,0,0)}.
\end{equation}
\par IoT devices are assumed to be static and distributed according to an
independent planar Poisson point process with intensity $ \lambda $ on a strip.
The finite transmission range of IoT devices is captured by having
the Poisson IoT devices restricted to a strip of width $ l $ centered at the $ x-$axis.
The proposed model allows us to specify the properties of uplink transmissions
from IoT devices to UAVs as the latter progress. Fig. \ref{fig:dronesandiots}
illustrates the UAVs, their projections onto the plane, and the IoT devices.
Thanks to the random shift $ U, $ the proposed UAV point process is stationary
(w.r.t. shifts along the $x$-axis)
and even jointly stationary with the device point process \cite{baccelli2010stochastic}.

\begin{remark}
When the UAVs are assumed to move in the positive direction of the $ x-$axis with speed $ v, $
the UAV point process at time $ t>0 $ is the initial UAV point process at time $ 0 $,
shifted by a vector $ (vt,0,0). $ The proposed UAV point process is translation-invariant
w.r.t. motion along the $ x$-axis, i.e., the distributions of the UAV point process
and of its shifted version are the same: $ \Phi(t) \stackrel{d}{=} \Phi(0). $
The translation-invariant structure provides the mathematical justification for
the definition of the typical UAV thanks to Palm calculus. 
The statistics of coverage and rate of the typical UAV will be evaluated in Section \ref{S:3}. 
\end{remark}

\subsection{Uplink Transmissions Associated With Motion}
In the proposed network, uplink transmissions from IoT devices are triggered by the motion of UAVs.
We assume that an IoT device is marked as \emph{active} if and only if it is located inside
the coverage area of an UAV, which will be referred to as its activation windows. 
The activation window is modeled by the set $ [-\frac{w}{2},\frac{w}{2}]\times[-\frac{l}{2},\frac{l}{2}] $
on the  plane, centered at the projection of every UAV onto the plane. 

\par The geometry of the activation window determines key feature of the proposed architecture,
and in particular the period of the activation process and the properties of the associated
uplink transmissions. As the UAV fleet moves in the positive $ x $-direction, the activation
windows also move. As a result, UAVs are able to harvest the data from all IoT devices on the
strip. This is what we called universal service above. The windows at time $ t $ are  
	\begin{align*}
	\cW(t)&=\bigcup_{i\in\bZ}\cW_i(t)\\
	&=\bigcup\limits_{i\in\bZ}\left[\mu i+vt -\frac{w}{2},\mu i+ vt+ \frac{w}{2}\right]\times \left[-\frac{l}{2},\frac{l}{2}\right].
	\end{align*} 
\par
\begin{figure}
	\centering
	\includegraphics[width=1\linewidth]{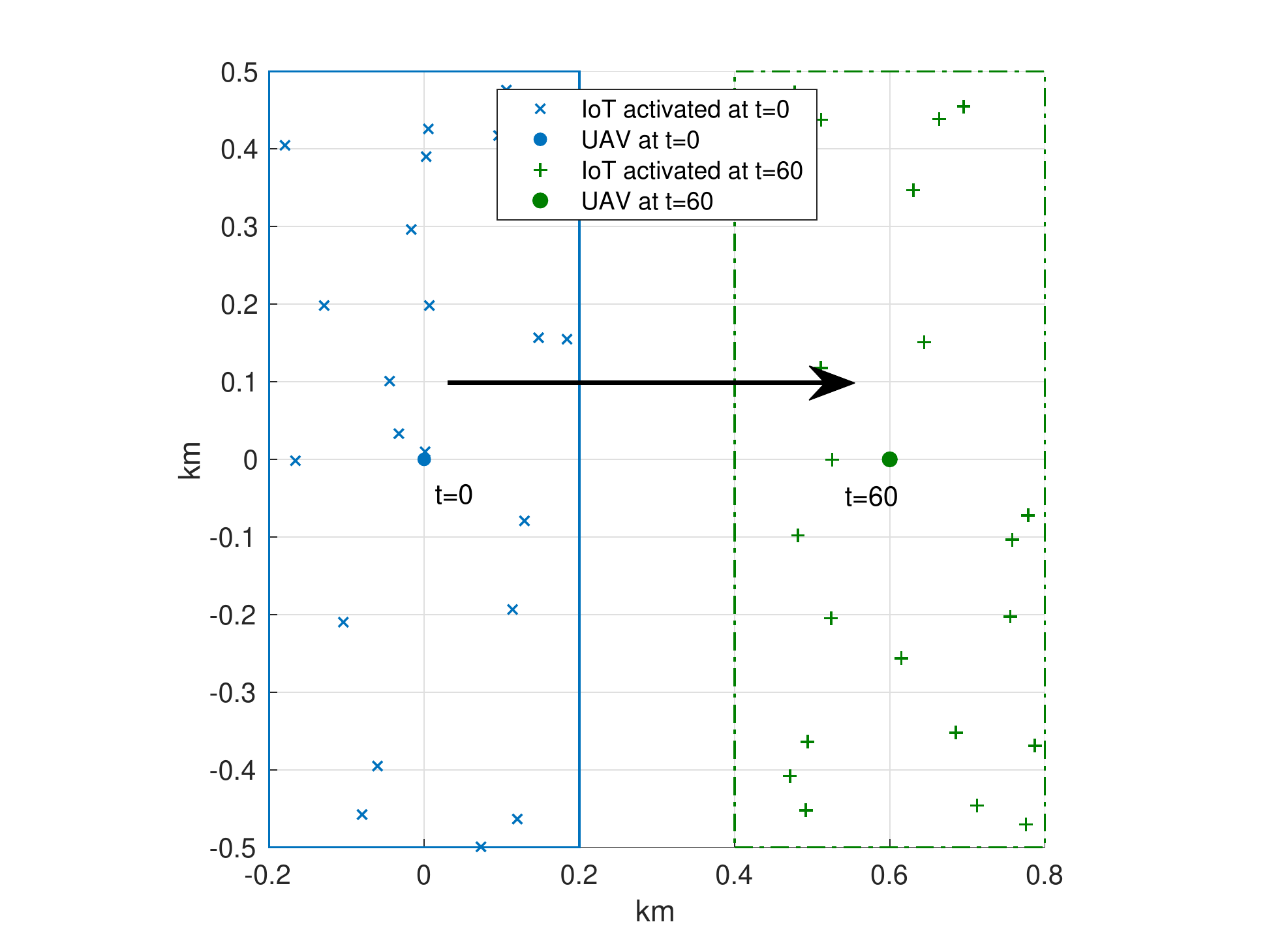}
	\caption{Illustration of the proposed network with a UAV located at $ (0,0,h) $ at time zero.
The black arrow indicates the direction of UAV motion. Each UAV moves at $ 36 $ km/h.
The density of IoT devices is $ 100/\text{km}^2 $. The activation windows at time
$ t=0, 60 $ are indicated by the solid blue and dotted green rectangles, respectively.
At each time slot, only one IoT device can upload its data to the corresponding UAV.
The access-granted IoT device at a certain time slot is uniformly selected in every window.}
	\label{fig:tdmapolicy}
\end{figure}

 To analyze the network performance, time is assumed to be slotted.
We assume that at each time slot, a single IoT device is randomly chosen from 
each window for the uplink transmission (we recall that windows do not intersect at
any given time). Hence, at each time, all IoT devices inside the window of an UAV are
activated as possible candidates for uplink transmission, and then at most one
IoT device is granted access for uplink transmission. Such an access control model
can be modeled by time division multiple access (TDMA) scheduling per UAV window,
which will later be approximated by a processor sharing scheme, which is equivalent to 
assuming that the time slot is very small compared to the time an IoT device spends
in the window of a passing by UAV. The TDMA scheduling is assumed to cope with the
fact that the transmit power of uplink transmitters is limited and the
network performance is interference limited.  Fig. \ref{fig:tdmapolicy} illustrates
the locations of IoT devices, associated activation windows at time $ 0 $ and $ 20 $,
and the corresponding active IoT devices, respectively.

\subsection{Propagation Model}\label{S:2C}
Time is slotted and each time slot has a duration $ T_s. $
Each time slot is assumed to be equal to the coherence time $ T_c $ of the wireless channel.
In other words, we assume that each slot experiences a different realization of a random fading.

To analyze the network performance, the received signal power is modeled by a
distance-based power law path loss model with Nakagami-m fading with parameters $ m $ and $ \Omega $. To maintain the analytical tractability of the paper, we do not consider air to ground channel model with direct line of sight component\footnote{The performance with line of sight channel component can be derived, leveraging the analysis presented in this paper.}.
The Nakagami-m fading is generally used to model various fading conditions including Rayleigh.
For instance, if $ m=\Omega=1, $ Nakagami-m fading boils down to Rayleigh.
The received signal power at distance $ d $ is given by $ pGd^{-\alpha} $,
where $ p$ is the transmit power, $ G $ is a Gamma random variable,
$G \sim \text{Gamma}(m, \Omega/m), $ and $ \alpha > 1$ is the path loss
exponent\cite{doi:10.1080/15326349.2014.990980}. 

\begin{table}\caption{Network Model and Parameters}
	\centering
\begin{tabular}{|c|c|}
		\hline
		Network parameters & Description\\
		\hline
		IoT devices & $ \Phi\sim $ Planar Poisson $ \lambda $ on the strip\\
		\hline 
		UAVs & Shifted comb process of inter distance $ \mu $\\
		\hline 
		UAV altitude & $ h $ \\ 
		\hline 
		UAV activation window & $ w \times l $ on the plane\\ 
		\hline 
		Multiple access & Time-division multiple access\\
		\hline
		UAV motion and speed &  $ x $-axis with speed $ v $\\ 
		\hline 
		Path loss exponent & $ \alpha >1$ \\ 
		\hline 
		UAV-IoT fading  & Gamma $ (m,\Omega/m) $\\ 
\hline 
	\end{tabular} 
\end{table}

\subsection{Performance Metrics}
For the proposed UAV network, we evaluate the performance from two typical perspectives:
the UAV's perspective and the IoT device's perspectives. Under the Palm distribution of
the UAV point process, we derive the coverage probability: $ \bP(\SIR>\tau) $ and the distribution of the
corresponding instantaneous data rate. Under the Palm distribution of the IoT point process,
we derive the amount of data uploaded from a typical IoT device to a UAV.

\section{Performance From the UAVs' Perspective}\label{S:3}
\subsection{Shot-noise Seen by a Typical UAV}
The shot-noise process $ N(t) $ is a stochastic process whose value at time $ t $
is the sum of the signal power received by the typical UAV from all the
access-granted IoT devices at time $ t $.
Due to the randomness of the locations of nodes and that of the wireless channel,
$N(t)$ is a random variable that will be characterized by its Laplace transform,
$ \bE[\exp(-sN(t))] $ with $ s $ the Laplace argument. The Laplace transform of
the shot-noise process has been extensively studied to obtain the key performance
metrics such as the coverage probability. In particular, the shot-noise process of
the Poisson point process has been used extensively \cite{baccelli2010stochastic}.
In this paper, the underlying IoT devices are modeled as a realization of the Poisson point process.
However, due to the TDMA processing at each UAV window and the movement of UAVs,
the access-granted IoT devices (which determine the shot-noise process) are not distributed
according to a Poisson point process. We hence need a specific analysis that is described below.

Consider the Palm distribution of the UAV point process. Without loss of generality,
under this Palm distribution, the typical UAV is located at $ (0,0,h) $ at a given
time slot \cite{baccelli2013elements}. Consequently, the activation windows for all UAVs are 
\begin{align*}
\cW=&\bigcup_{i\in\bZ}\cW_i=\bigcup\limits_{i\in\bZ}\left[\mu i-\frac{w}{2},\mu i+ \frac{w}{2}\right]\times \left[-\frac{l}{2},\frac{l}{2}\right].
\end{align*} 
Under the TDMA scheduling, the shot-noise process seen at the typical UAV is given by 
\begin{align}\label{3}
N=\sum_{(X_i,Y_i) \in\hat{\Phi}} p G_i  \|(X_i,Y_i,0)-(0,0,h)\|^{-\alpha}\mathbbm{1}_{\{\Phi(\cW_i)\neq \emptyset\}},
\end{align}
where $ (X_i,Y_i) $ are the $ x $ and $ y $ coordinates of the transmitting IoT device,
if any, in window $ \cW_i $. The point process $ \hat{\Phi} $ denotes the access-granted
IoT point process. Let $ \mathbbm{1}_{\{A\}}  $ denote the indicator function that takes
value one if $ A $ is true, or zero otherwise.

	\begin{figure*}

\end{figure*}
\begin{theorem}\label{T:1}
$ \cL_{N}(s) $ the Laplace transform of the uplink shot-noise process at the typical UAV is given by Eq. \eqref{eq:L1} where $ s $ is the Laplace transform argument. 
\end{theorem}
\begin{IEEEproof} 
We can write $ \cL_N(s) $ as follows:
 	\begin{align}
	&\bE_{\Psi}^0\left[e^{-s\sum_{(X_i,Y_i)\in\cW_i}^{i\neq 0} p G_{i}{\|(X_i,Y_i,0)-(0,0,h)\|}^{-\alpha}\ind_{\{\Phi(\cW_i)\neq \emptyset\}}}\right]\nnnl
	&\ea\bE_{\Psi}^0\left[ \prod_{X_i,Y_i} \bE\left[ \left. e^{{-s p G_{i}\|(X_i,Y_i,-h)\|^{-\alpha}\ind_{\{\phi(\cW_i)\neq \emptyset\}}}}\right.\right]\right],
	\end{align}
To derive (a), we use the fact that the points from disjoint windows are independent.
Using that the number of points in each window follows the Poisson distribution, we have 
	\begin{align}
	&\cL_N(s)\nnb\\
	&=\bE_{\Psi}^0\left[\left.\prod_{i\in\bZ}\left(\bP(\Phi(\cW_i)=\emptyset)\right.\right.\right.\nnb\\
	&\hspace{14mm}\left.\left.\left.+\bP(\Phi(\cW_i)\neq \emptyset)\cdot\bE\left[\left.e^{-s p G_{i}\|(X_i,Y_i,-h)\|^{-\alpha}}\right.\right]\right.\right)\right]\nnnl
	&\eb\bE_{\Psi}^0\left[\prod_{i\in\bZ}\left(e^{-\lambda w l}\right.\right.\nnb\\
	&\hspace{14mm}\left.\left.+\left({1-e^{-\lambda w l}}\right)\int_{\text{supp}(G)}\hspace{-7mm}e^{-spg{(X_i^2+Y_i^2+h^2)}^{-\frac{\alpha}{2}}}f_G(g)\diff g\!\!\right)\!\right]\nnb\\
	&\ec\prod_{i\in\bZ}\left(e^{-\lambda w l}+\frac{1-e^{-\lambda w l}}{ w l}\cdot\right.\nnb\\
	&\left.\hspace{10mm}\int_{i\mu-\frac{w}{2}}^{i\mu+\frac{w}{2}}\int_{-\frac{l}{2}}^{\frac{l}{2}}\int_{\text{supp}(G)}\hspace{-7mm}e^{-spg{(x^2+y^2+h^2)}^{-\frac{\alpha}{2}}}f_G(g)\diff g\diff y \diff x\right)\nnb\\
	&\ed\prod_{i\in\bZ}\left(e^{-\lambda w l}+\frac{1-e^{-\lambda w l}}{ w l}\cdot\right.\nnb\\
	&\left.\hspace{10mm}\int_{i\mu-\frac{w}{2}}^{i\mu+\frac{w}{2}}\int_{-\frac{l}{2}}^{\frac{l}{2}}\left(\frac{1}{1+\frac{sp\Omega m^{-1}}{{(x^2+y^2+h^2)}^{\frac{\alpha}{2}}}}\right)^{m}\!\!\!\diff y \diff x\right).\nnb
	\end{align}
To derive (b), we use the fact that the probability that the window $ \cW_i $
is empty of points is equal to $  \exp(-\lambda wl).$
To derive (c), we use the fact that, given that the window is not empty of points,
a randomly selected point is uniformly distributed inside the window
$ \cW_i = [i \mu-\frac{w}{2} ,i \mu +\frac{w}{2}]\times [-\frac{l}{2},\frac{l}{2}] $,
and the selected points across windows are also independent.
To get (d), we use the Laplace transform of the exponential random variable with mean one.
\end{IEEEproof}
	
Note that the integral formula, Eq. \eqref{eq:L1}, is not equal to zero because (1)
the shot-noise process $ N $ is statistically dominated by the shot-noise process 
seen from the origin and (2) the Laplace transform of the planar Poisson shot-noise
process is not equal to zero. See  \cite{baccelli2010stochastic,haenggi2009interference}
for the Laplace transform of the planar Poisson shot-noise process. 
	\par
\begin{figure}
	\centering
	\includegraphics[width=1\linewidth]{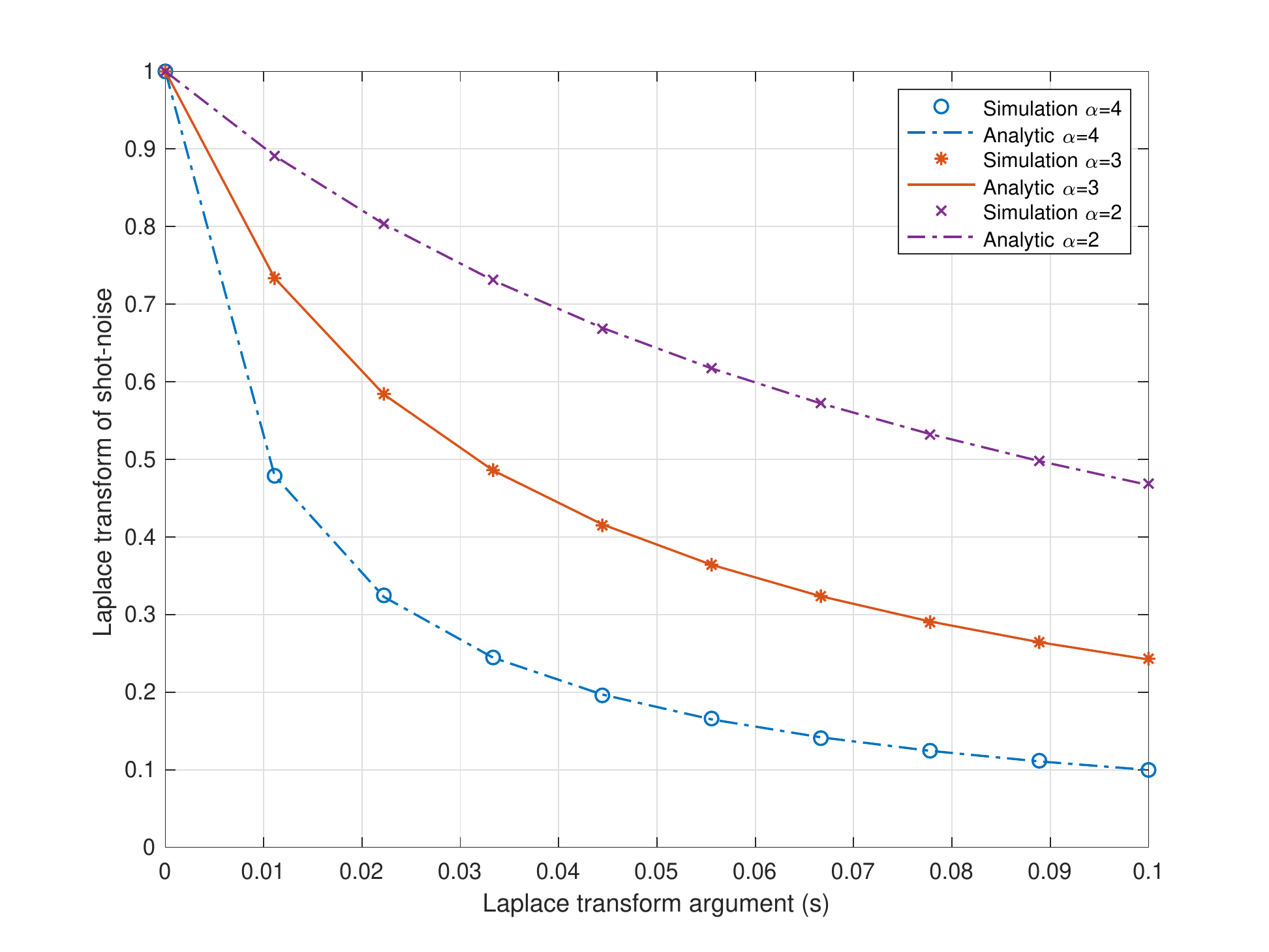}
	\caption{Laplace transform of the shot-noise process seen by the typical UAV with Nakagami-m fading parameters $ m=\Omega=1 $. We consider $ \lambda = 1000/\text{km}^2 $, $ \mu=2  $ km, $ w=0.25 $ km, $ l=0.5 $ km, and $ h= 0.25 $ km. Simulation results exactly match the analytically derived formula.}
	\label{fig:shotnoiselaplace}
\end{figure}

 
Fig. \ref{fig:shotnoiselaplace} illustrates the shot-noise process obtained from the
derived formula and from Monte Carlo simulations, respectively. It confirms that the
derived formula is accurate. The figure shows that the Laplace transform decreases
as the path loss exponent increases. 
\begin{remark}
The shot-noise process of the typical UAV is time-invariant. Note that the time-invariance
property of the shot-noise process comes from the fact that the underlying IoT point process
is modeled by a planar Poisson point process and the fact that the UAV point process is
translation and time-invariant. The shot-noise process determines the interference 
seen by the typical UAV (defined as the sum of the powers of all signals
received by the typical UAV except that of the IoT device that transmits to it at that time),
which in turn determines the SIR or SINR and the link performance. 
\end{remark}

\begin{corollary}
$ \cL_{I}(s) $ the Laplace transform of the interference seen by the typical UAV is given by Eq. \eqref{L:I}.
\end{corollary}
The activation window $ [-w/2,w/2]\times [-l/2,l/2] $ is used to schedule the transmissions of IoT devices.
For the evaluation of interference, the signal component from the window $ \cW_0 $ is removed from
Eq. \eqref{T:1}.
\begin{example}
For the parameters $ m=\Omega=1, $ Nakagami fading boils down to Rayleigh.
In this case, $ \cL_{\bar{I}}(s) $ the Laplace transform of the shot-noise is given by Eq. \eqref{L:I:R} 
\end{example}
\begin{figure*}
		\begin{equation}\label{eq:L1}
	\prod_{i \in\bZ}\left(e^{-\lambda w l}+\frac{1-e^{-\lambda w l }}{wl}\int_{i\mu-\frac{w}{2}}^{i\mu+\frac{w}{2}}\int_{-\frac{l}{2}}^{\frac{l}{2}}\left(\frac{1}{1+\frac{sp\Omega m^{-1}}{{(x^2+y^2+h^2)}^{{\alpha}/{2}}}}\right)^m\diff y \diff x\right).
	\end{equation}
		\begin{align}
	&\prod_{i\in\bZ}^{\neq 0}\left(e^{-\lambda w l}+\frac{1-e^{-\lambda w l}}{ w l}\int_{i\mu-\frac{w}{2}}^{i\mu+\frac{w}{2}}\int_{-\frac{l}{2}}^{\frac{l}{2}}\left(\frac{1}{1+sp\Omega m^{-1}{(x^2+y^2+h^2)}^{-\frac{\alpha}{2}}}\right)^{m}\!\!\!\diff y \diff x\right).\label{L:I}
	\end{align}
		\begin{align}
	\prod_{i\in\bZ}^{}\left(e^{-\lambda w l}+\frac{1-e^{-\lambda w l}}{ w l}\int_{i\mu-\frac{w}{2}}^{i\mu+\frac{w}{2}}\int_{-\frac{l}{2}}^{\frac{l}{2}}\frac{1}{1+sp{(x^2+y^2+h^2)}^{-\frac{\alpha}{2}}}\diff y \diff x\right).\label{L:I:R}
	\end{align}
			\rule{\textwidth}{0.2pt}	\vspace{-1em}
\end{figure*}
\subsection{Coverage Probability of the Typical UAV}
We assume that each UAV decodes the received signal transmitted from its own window,
by treating interference from other windows as noise.
The coverage probability is defined as a function of $ \tau $: $ \bP_{\Psi}^0(\SIR\geq\tau) $
under the Palm distribution of $ \Psi $, where $ \SIR=S/(I+n) $ and the random variables $ S $ and 
$ I $ denote the received signal power and the received interference power, respectively. 

\begin{figure}
	\centering
	\includegraphics[width=1\linewidth]{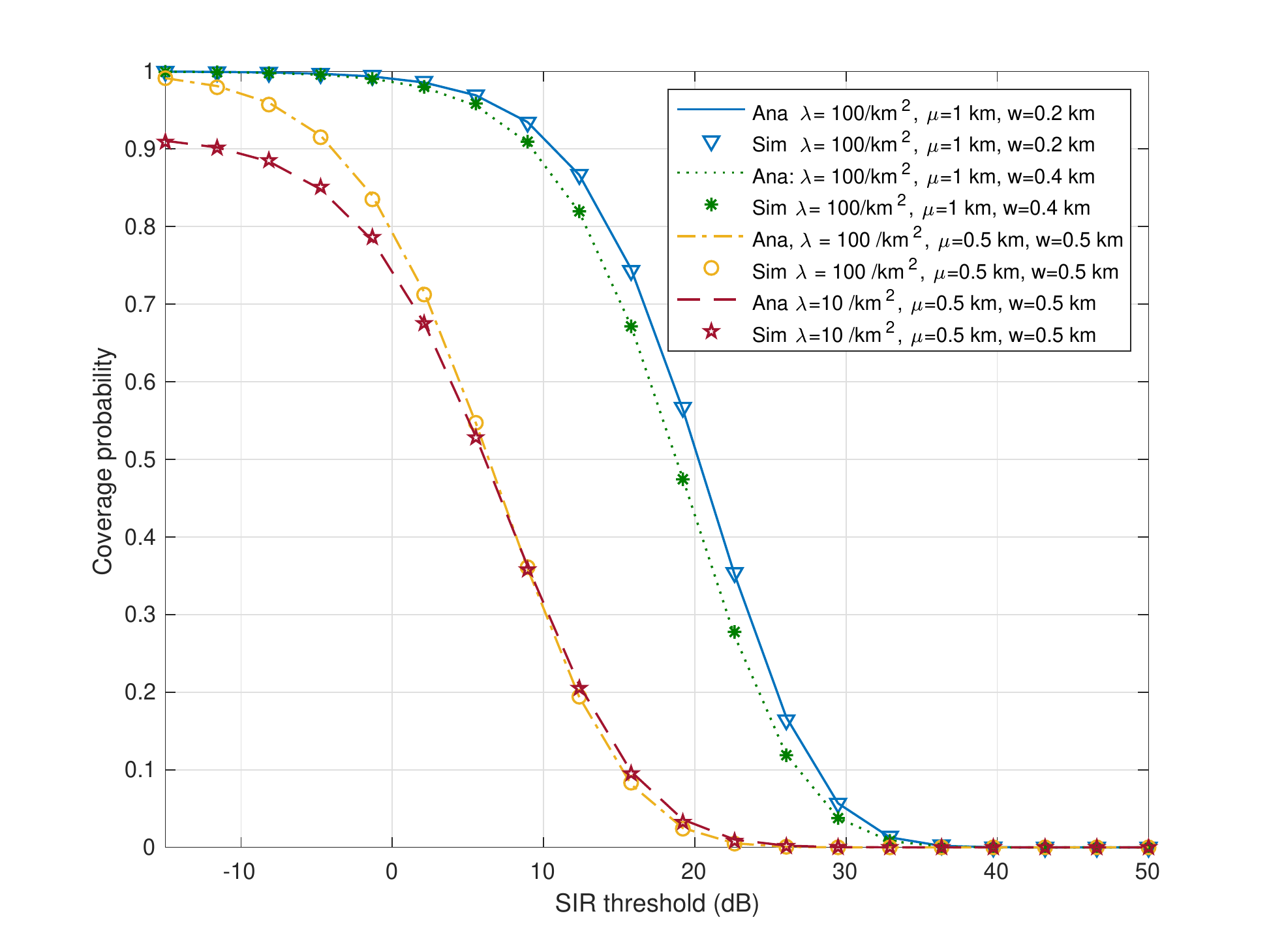}
	\caption{The coverage probability of the typical UAV. We consider $ \alpha=4 $, $ h= 0.2 $ km, $ l=0.5 $ km, and $\Omega=m=1 $.  }
	\label{fig:coverageprobabilitymatch}
\end{figure}

\begin{theorem}\label{T:2}
The coverage probability of the typical UAV is 
	\begin{equation}\label{eq:T1}
\frac{1-e^{-\lambda w l}}{ w l}\int_{-\frac{w}{2}}^{\frac{w}{2}}\int_{-\frac{l}{2}}^{\frac{l}{2}} \left.\left.\uta{\sum_{i=0}^{m-1}\frac{(-s)^{i}}{i!}\frac{\diff^i}{\diff s^i}\cL_{I}(s)}\right.\right.\diff y \diff x,
	\end{equation}
where term (a) being evaluated at $ s=\frac{\tau m (x^2+y^2+h^2)^\frac{\alpha}{2}}{p\Omega} $ and $ \cL_I(s) $ is the Laplace transform of the interference at the typical UAV provided in Eq. \eqref{L:I}.
\end{theorem}
\begin{IEEEproof}
First, the coverage probability of the typical UAV is derived under the Palm distribution
of the UAV point process, as in the derivation of the shot-noise process.
The coverage probability of the typical UAV is given by
	\begin{align}
		&\bP_{\Psi}^0(\SIR\geq\tau)\nnb\\
		&={\bP_{\Psi}^0(\SIR\geq\tau|\Phi(\cW_0)\neq \emptyset)}\bP(\Phi(\cW_0)\neq \emptyset)\nnb\\
		&\hspace{3mm}+\uta{\bP_{\Psi}^0(\SIR\geq\tau|\Phi(\cW_0)= \emptyset)}\times \bP_{\Psi}^0(\Phi(\cW_0)= \emptyset)\nnb\\
		&={\bP_{\Psi}^0(\SIR\geq\tau|\Phi(\cW_0)\neq \emptyset)}\bP(\Phi(\cW_0)\neq \emptyset)\nnb,
	\end{align}
where we use the fact that the coverage probability is considered to be zero if
the window $ \cW_0 $ is empty of points; term (a) vanishes. Let $ X_0 $ and $ Y_0 $ 
denote the $ x,y $ coordinates of the transmitting IoT device in the window of
the typical UAV, $ \cW_0 $, respectively. Then, we have 
		\begin{align*}
		&\bP(\SIR\geq\tau|\Phi(\cW_0)\neq \emptyset)\\
		&=\bP_{\Psi}^0\!\left(\!\frac{pG{\|(X_0,Y_0,-h)\|}^{-\alpha}}{\sum\limits_{(X_i,Y_i)\in{\Phi(\cW_i)}}^{\setminus(X_0,\!Y_0)}\!\!\!\!\!\!pG{\|(X_i,Y_i,-h)\|}^{-\alpha}\mathbbm{1}_{\{\Phi(\cW_i)\neq \emptyset\}}}\!\geq\!\tau\right)\\
		&\eb\bP_{\Psi}^0\left(G\geq{I\tau p^{-1}{(X_0^2+Y_0^2+h^2)}^{\frac{\alpha}{2}}}\right)\\
		&=\bE_{X,Y,I}\left[\left.\sum_{i=0}^{m-1}\frac{(-s)^{i}}{i!}\frac{\diff^i}{\diff s^i}e^{-ms/\Omega}\right|_{s=I\tau p^{-1}(X^2+Y^2+h^2)^\frac{\alpha}{2}}\right],
	\end{align*}
where $ \bE_{X_0,Y_0,I} $ denotes the expectation w.r.t. the random variables $ X_0, Y_0 $ and $ I, $
where $ I $ is the interference seen by the typical UAV at the origin.
Given that the location of the access-granted IoT device in the typical window $ (X_0, Y_0) $
is independent of the interference, the coverage probability of the typical UAV is given by 
	\begin{align*}
&\frac{\bP_{\Psi}^0(\Phi(\cW_0)\neq \emptyset)}{wl}\nnb\\
		&\times\int_{-\frac{w}{2}}^{\frac{w}{2}}\!\!\int_{-\frac{l}{2}}^{\frac{l}{2}}\!\!\left(\left.\sum_{i=0}^{m-1}\frac{(-s)^{i}}{i!}\frac{\diff^i}{\diff s^i}\cL_{I}(s)\right|_{s=\frac{\tau m(x^2+y^2+h^2)^\frac{\alpha}{2}}{p\Omega }}\right)\!\!\diff y \diff x\nnb,
	\end{align*}
where the Laplace transform of the interference $ \cL_{I}(s) $ is given in Eq. \eqref{L:I}.  
\end{IEEEproof}
\begin{remark}
Since the UAV and IoT point processes are jointly stationary and the IoT point process is mixing,
the above coverage expression can be interpreted in an ergodic sense \cite{baccelli2013elements}.
The coverage probability is also the proportion of time that the typical UAV has an IoT device
in its window and successfully decodes its signal.
\end{remark}
	\begin{figure*}
	\begin{equation}
	\frac{1-e^{-\lambda w l}}{ w l}\!\!\!\int_{-\frac{w}{2}}^{\frac{w}{2}}\int_{-\frac{l}{2}}^{\frac{l}{2}}\prod_{k\in\bZ\setminus 0}\left(e^{-\lambda wl }+\frac{1-e^{-\lambda  w l}}{ w l}\int_{k\mu-\frac{w}{2}}^{k\mu+\frac{w}{2}}\!\!\!\int_{-\frac{l}{2}}^{\frac{l}{2}}\frac{\diff u \diff v}{\left.1+\frac{\tau{(x^2+y^2+h^2)}^{\frac{\alpha}{2}}}{{(u^2+v^2+h^2)}^{\frac{\alpha}{2}}}\right.}\right)\diff y\diff x.\label{eq:T1-1}
	\end{equation}
		\rule{\textwidth}{0.2pt}	\vspace{-1em}
\end{figure*}
\begin{example}
	With fading parameters $ m=\Omega=1, $ the coverage probability is given by Eq. \eqref{eq:T1-1}.
\end{example}
Fig. \ref{fig:coverageprobabilitymatch} illustrates the coverage probability of the typical UAV
obtained by formula Eq. \eqref{eq:T1-1}. Simulation results are presented to validate the accuracy 
of the derived results. The simulation results and analytical results match. We found that when
$ \lambda = 100 /\text{km}^2$, and $ \mu=1 $ km, an increase of the window size decreases
the SIR coverage probability of the typical UAV due to the decrease of the desired signal power
and the increase of interference. In addition, there exists an upper bound on the coverage probability,
which is given by $ 1-e^{-\lambda wl} $. This value corresponds to the probability that the 
window of the typical UAV is not empty of IoT devices. In this sense, the coverage probability
captures the utilization at UAVs.

\begin{figure}
	\centering
	\includegraphics[width=1\linewidth]{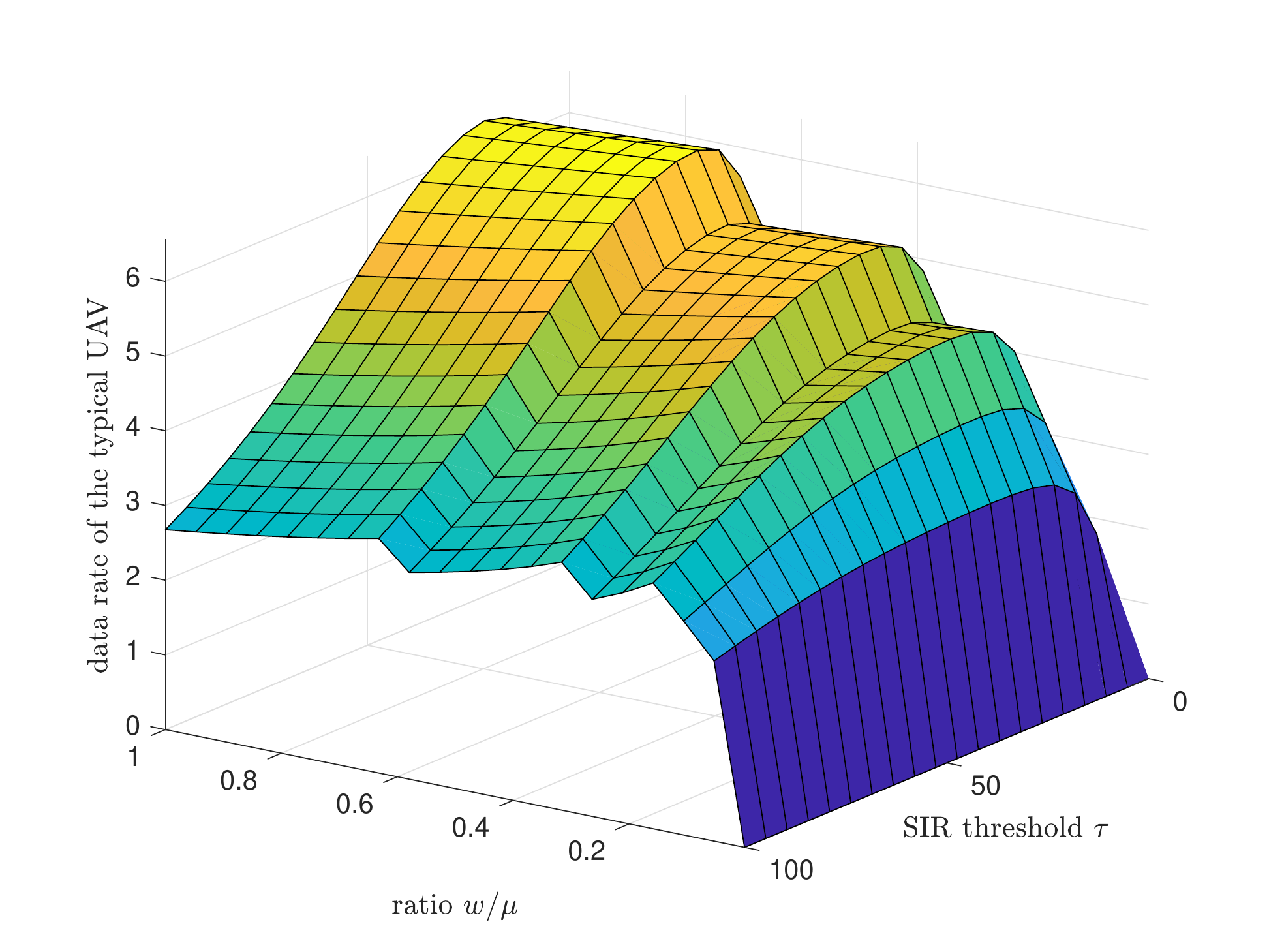}
	\caption{The data rate at the typical UAV where a typical $M$-ary modulation: $M= 2^{\floor{\log_2(1+\tau)}}, \Omega= m=1 $. Note that the $ z $-axis is in bit/sec/Hz.}
	\label{fig:dataratesirthresholddensityfloormodulation}
\end{figure}


\begin{figure}
	\centering
	\includegraphics[width=1\linewidth]{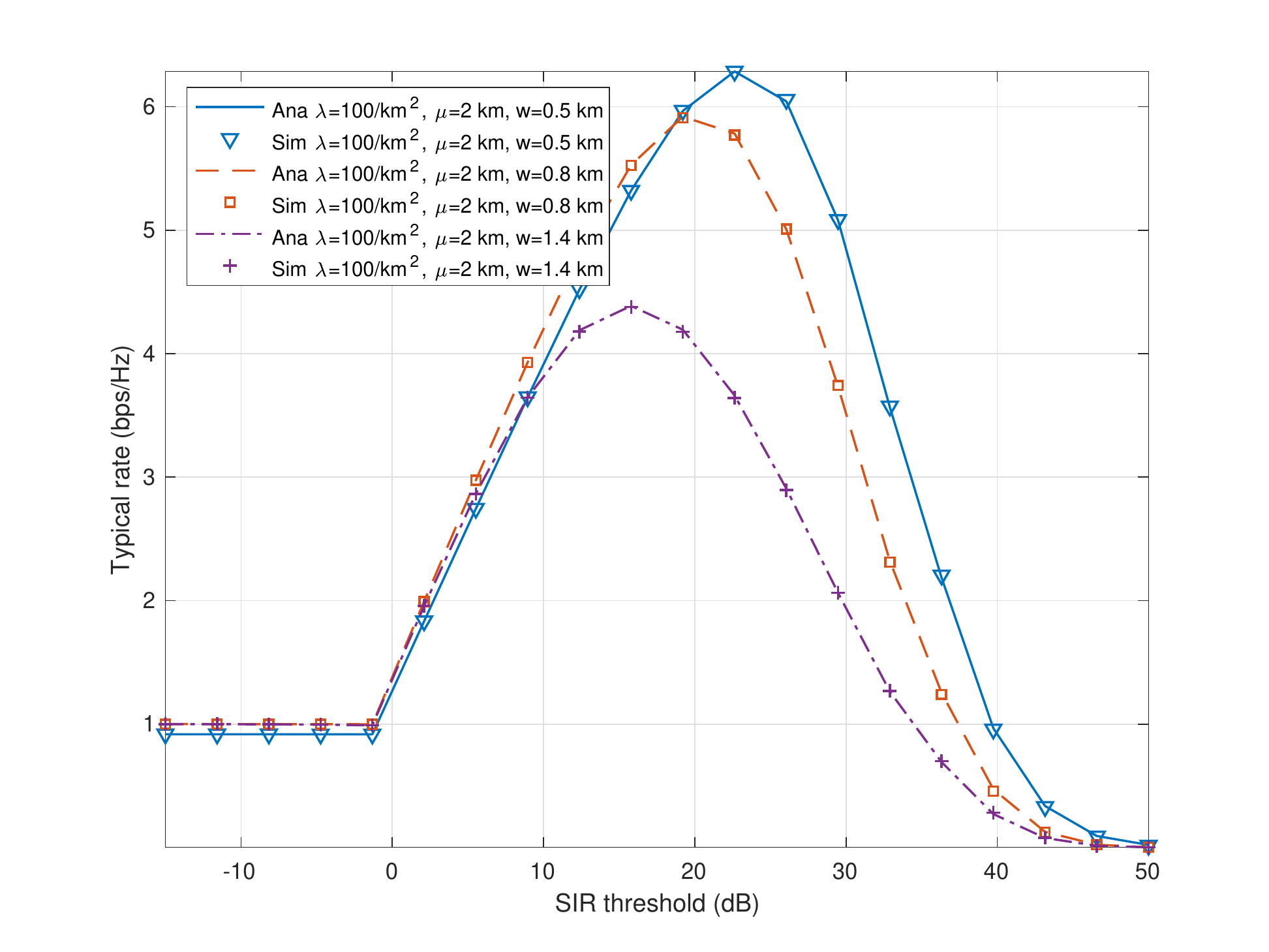}
	\caption{The data rate seen by the typical UAV.
We consider $ \alpha=4,$ $ h= 0.2 $ km, $ l=0.5 $ km and $ m=\Omega = 1. $
The SIR threshold on the $ x $-axis indicates the minimum SIR value for  
which uplink communications are reliable.}
	\label{fig:shannonratetau}
\end{figure}

\subsection{Mean Data Rate of the Typical UAV}
This section analyzes the mean UAV data rate and the associated spectral efficiency.
Let us consider a fixed $ M $-ary modulation with minimum SIR threshold $ \tau $.
The mean data rate of the typical UAV is then defined by $ \log_2(M)\bP_{\Psi}^0(\SIR\geq \tau) $.
\begin{remark}
{For practical reasons, we consider a fixed $ M$-ary  modulation at the IoT devices transmitters.
Under the TDMA scheduling where the transmission channel from the IoT devices to the UAV varies 
w.r.t. space and time, channel-side information is unavailable at the UAVs. Consequently, the
fixed modulation with threshold $ \tau $ enables reliable uplink transmissions and
reduces the chance of transmission failure.}
\end{remark}
	\begin{figure*}
	\begin{align}
	\frac{\log_2(M)(1-e^{-\lambda w l})}{ w l}\int_{-\frac{w}{2}}^{\frac{w}{2}}\int_{-\frac{l}{2}}^{\frac{l}{2}} \left(\left.\sum_{i=0}^{m-1}\frac{(-s)^{i}}{i!}\frac{\diff^i}{\diff s^i}\cL_{I}(s)\right|_{s=\frac{\tau m (x^2+y^2+h^2)^\frac{\alpha}{2}}{p\Omega}}\right)\diff y \diff x.\label{eq:r}
	\end{align}
			\begin{equation}\label{eq:par}
	\frac{\log_2(M)(1-e^{-\lambda w l)}}{ w l}\!\!\!\int_{-\frac{w}{2}}^{\frac{w}{2}}\int_{-\frac{l}{2}}^{\frac{l}{2}}\prod_{k\in\bZ\setminus 0}\left(e^{-\lambda wl }+\frac{1-e^{-\lambda  w l}}{ w l}\int_{k\mu-\frac{w}{2}}^{k\mu+\frac{w}{2}}\!\!\!\int_{-\frac{l}{2}}^{\frac{l}{2}}\frac{\diff u \diff v}{\left.1+\frac{\tau{(x^2+y^2+h^2)}^{\frac{\alpha}{2}}}{{(u^2+v^2+h^2)}^{\frac{\alpha}{2}}}\right.}\right)\diff y\diff x.
	\end{equation}
			\begin{equation}
	\frac{(1-e^{-\lambda wl})\log_2(M)}{\lambda wl^2}\frac{1}{v}\int_{-\frac{w}{2}}^{\frac{w}{2}}\int_{-\frac{l}{2}}^{\frac{l}{2}}\left.\sum_{i=0}^{m-1}\frac{(-s)^{i}}{i!}\frac{\diff^i}{\diff s^i}\cL_{I}(s)\right|_{s=\frac{\tau m  (x^2+y^2+h^2)^\frac{\alpha}{2}}{p\Omega}}\diff y \diff x\label{eq:T3}.
	\end{equation}
			\rule{\textwidth}{0.2pt}	\vspace{-1em}
\end{figure*}
\begin{theorem}\label{T:1-rate}
The mean {data rate seen by the typical UAV}, $ \cR, $ is given by Eq. \eqref{eq:r}.

\end{theorem}
\begin{IEEEproof}
The proof immediately follows from Theorem \ref{T:2}.
\end{IEEEproof}

\begin{example}
With fading parameters $ m=\Omega=1, $ the typical data rate is given by Eq. \eqref{eq:par}

\end{example}

Figs. \ref{fig:dataratesirthresholddensityfloormodulation} and \ref{fig:shannonratetau}
plot the mean uplink data rate seen by the typical UAV.
The surface in Fig. \ref{fig:dataratesirthresholddensityfloormodulation} is
non-smooth because of the condition $ M=2^{\floor{\log_2(1+\tau)}} $.
Fig. \ref{fig:shannonratetau} illustrates the mean data rate of the typical UAV
for various SIR thresholds $ \tau.  $ It shows that the analytically derived data rate results
exactly match the simulated data rate results. The mean data rate is given by 
$ \log_2(1+\tau) \bP(\SIR\geq \tau) $,
which is the product of an increasing and a decreasing function w.r.t. $ \tau  $.
This explains why, for a given density of IoT devices and UAVs, there exists an optimal
SIR threshold $ \tau $ that maximizes the data rate seen by the typical UAV.
Notice that the SIR threshold may not be as configurable as other network parameters, like, e.g.,
the scheduling window $ w $ because, in practice, the value might be determined by 
the transmit power of the IoT devices and the bandwidth that these devices use.

\begin{remark}
Consider an IoT deployment scenario where the chance of having a window with no IoT device
is very small; $ \bP(\Phi(\cW_0)\neq\emptyset)\approxeq 1. $ Such a condition would be met
for parameters  $ \lambda = 500 /\text{km}^2, $ $ w= 100 $ m, and $ l=100 $ m.
Conditioning on the fact that the typical window $ \cW_0 $ is not empty of IoT devices,
the coverage probability of the typical UAV, $ \bP_{\Psi}^0(\SIR\geq\tau|\Phi(\cW_0)\neq \emptyset) $, is approximately given by 
	\begin{align}
	\frac{1}{ w l}\int_{-\frac{w}{2}}^{\frac{w}{2}}\int_{-\frac{l}{2}}^{\frac{l}{2}} \underbrace{\sum_{i=0}^{m-1}\frac{(-s)^{i}}{i!}\frac{\diff^i}{\diff s^i}\cL_{I}(s)}_{s=\frac{\tau m  (x^2+y^2+h^2)^\frac{\alpha}{2}}{p \Omega}}\diff y \diff x.\nnb
	\end{align}
Conditionally on the fact that its window is not empty of IoT devices,
the data rate $ \bar{\cR} $ of the typical UAV is now of the form 
	\begin{equation}
	\frac{\log_2(M)}{ w l}\int_{-\frac{w}{2}}^{\frac{w}{2}}\int_{-\frac{l}{2}}^{\frac{l}{2}} \left.\underbrace{\sum_{i=0}^{m-1}\frac{(-s)^{i}}{i!}\frac{\diff^i}{\diff s^i}\cL_{I}(s)}_{s=\frac{\tau m  (x^2+y^2+h^2)^\frac{\alpha}{2}}{p\Omega}}\right.\diff y \diff x.\label{cR}
	\end{equation}
The above quantity in Eq. \eqref{cR} should not be confused with $ \cR $,
the data rate of the typical UAV in Eq. \eqref{eq:r}. 
When the density of IoT devices is high enough to ensure that the window is rarely empty of IoT device,
$ \cR \approxeq \bar{\cR}. $
\end{remark}

\section{Performance From the IoT Devices' Perspective}\label{S:4}
This section focuses on the network performance as seen by a typical IoT device. 
Specifically, we derive the total amount of data transmitted, i.e., harvested from 
the typical IoT device to a UAV, while the former is inside the activation window of the latter.
This is evaluated under the Palm distribution of the IoT point process,

\subsection{Mean Amount of Data Transmitted from a Typical IoT Device}
The coverage probability and rate analysis in the previous section capture the 
instantaneous network performance at a UAV.
This subsection is focused on the evaluation of 
the meant amount of data transmitted from the typical IoT device per UAV passage. 

Here are a few preliminary observations and definitions before stating the main theorem.
We recall that in each activation window, there is a random (Poisson) number
of IoT devices present at each time slot and that the processing by the UAV 
is shared between these IoT devices, based on TDMA. 
As the UAV moves, the TDMA scheme involves a different set of devices.
Consequently, in order to compute the mean amount of data uploaded from a typical IoT device
to a UAV while it is inside the activation window of the UAV, 
one should consider the evolution of the network geometry w.r.t. time. 
Typicality, above and in the theorem below, is again defined in terms of the Palm probability
of the IoT point process.

\begin{figure}
	\centering
	\includegraphics[width=1\linewidth]{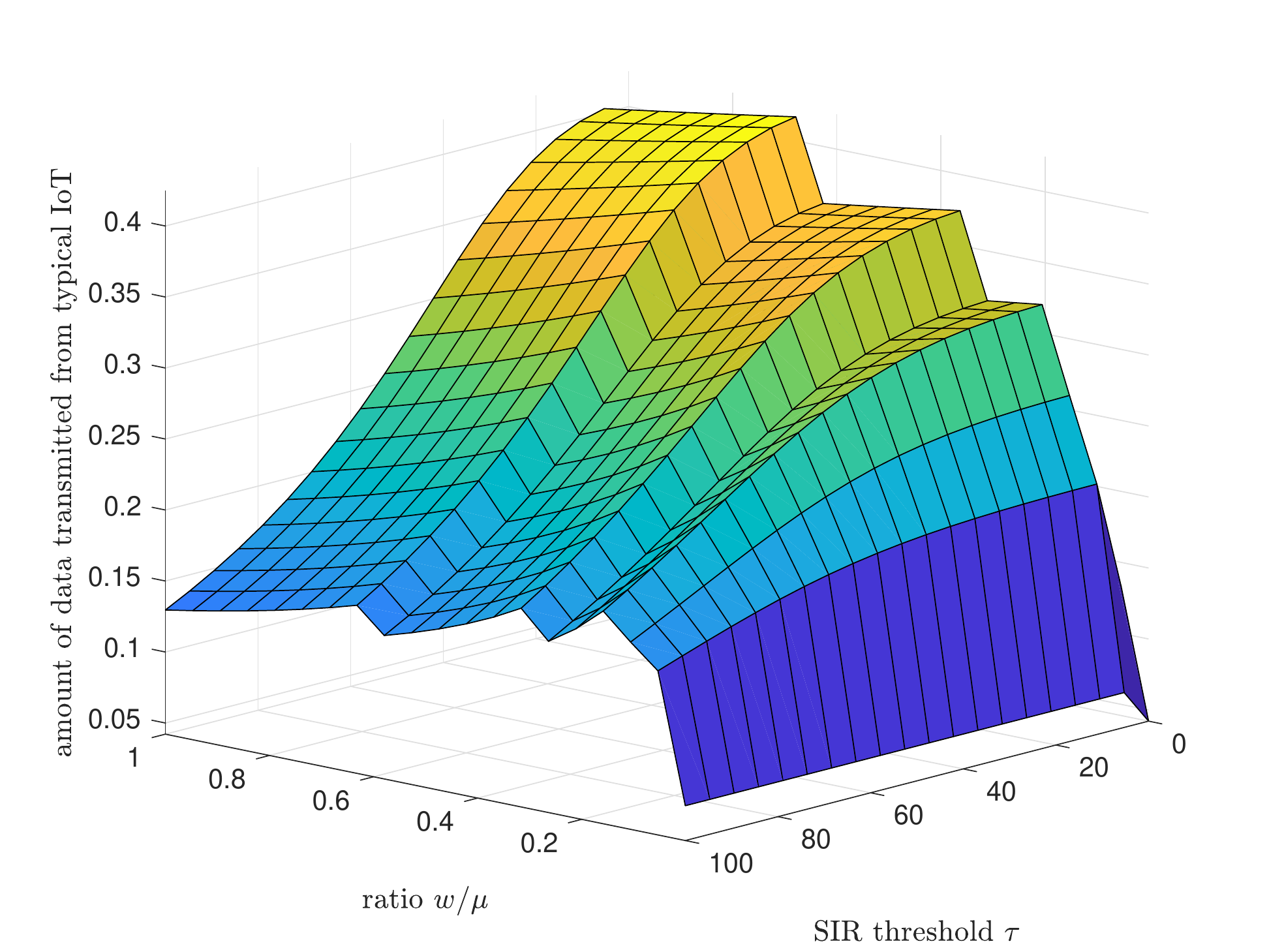}
	\caption{The amount of data transmitted from the typical IoT device to its serving UAV. We consider $ \alpha=3.5,$ $ h=0.2 \text{ km},$ $ \mu=2\text{ km}, $ $  l=0.5 \text{ km},$ $ v=30 \text{m/sec} $, Nakagami fading parameters $ \Omega=m=1, $ and IoT density $ \lambda=1000/\text{km}^2$. Note that the $ z $-axis is in bit/Hz. }
	\label{fig:amountofdatatransmitted}
\end{figure}

\begin{theorem}\label{T:3}
The mean amount of data $ \cD $ harvested from the typical IoT device per UAV passage is given by Eq. \eqref{eq:T3}. 
\end{theorem}
\begin{IEEEproof} The $ y $ coordinates of the points of the IoT point process
$ \Phi $ of intensity $ \lambda  $ on the strip can be considered as i.i.d. marks
of a linear Poisson point process $ \tilde{\Phi}  $ with intensity $ \lambda l $
on the $ x $-axis. Under the Palm distribution of the IoT point process,
the typical IoT device is located at the origin with its mark $ Y\sim \text{Uniform}[-\frac{l}{2},\frac{l}{2}] $. 
The total amount of data transmitted from the typical IoT device to its serving UAV is the given by 
	\begin{align}
		\cD&=\bE_{\tilde{\Phi}}^0\left[\sum_{k=-\floor{\frac{w}{2vT_s}}}^{\floor{\frac{w}{2vT_s}}}\cR(Y,kT_s,\tilde{\Phi})\right]\label{12},
	\end{align}
where $ \bE_{\tilde{\Phi}}^0 $ denotes the Palm expectation w.r.t. $ \tilde{\Phi} $,
$ \cR(Y,kT_s,\tilde{\Phi},\Psi(k)) $ is the amount of data that the typical IoT device,
with its mark randomly distributed between $ -\frac{l}{2} $ and $ \frac{l}{2} $,
transmits to its serving UAV at time slot $ k $.
The rate $ \cR(Y,kT_s,\tilde{\Phi}) $ is given by 
	\begin{align}
		\cR(Y,kT_s,\tilde{\Phi})&= T_s \log_2(M)\nnb\\
		&\times\ind_{\{\SIR_{(0,Y,0)\to(-vkT_s,0,h)\geq\tau}\}}\nnb\\
		&\times\ind_{\{(0,Y) \text{ is selected to transmit at slot $ k $}\}}.
	\end{align}
	The above  formula is justified when we assume that $ T_s $ is sufficiently small,
so that the relative positions of $(0,Y)$ and the position of the UAV $(-vkT_s,0,h)$ do not
vary much during the $k$-th time slot.  
Under this assumption, one can further approximate the summation of Eq. \eqref{12}
as the following integral:
		\begin{align}
	\cD&\approxeq\bE_{\tilde{\Phi}}^0\left[\int_{-\frac{w}{2v}}^{\frac{w}{2v}} \cR(Y,t,\tilde{\Phi}) \diff t \right],\label{17}
	\end{align}
	where the continuous rate $ \cR(Y,t,\tilde{\Phi}) $ is now given by 
	\begin{align}
		&\log_2(M)\ind_{\{\SIR_{(0,Y,0)\to(vt,0,h)\geq\tau}\}}\ind_{\{Y \text{ is selected to transmit at time $ t $}\}}\nnb\\
		&=\log_2(M)\!\sum_{k=1}^{\infty}\left(\!\!\ind_{\{\SIR_{(0,Y,0)\to(vt,0,h)\geq\tau}\}}\right.\nnb\\
		&\hspace{23mm}\left.\ind_{\{Y \text{ is selected out of $ k $}|\tilde{\Phi}{(\cW_{vt})=k}\}}\ind_{\{\tilde{\Phi}(\cW_{vt})=k\}}\!\right)\nnb\\
		&=\log_2(M)\sum_{k=1}^{\infty}\left(\ind_{\{Y \text{ is selected out of $ k $}|\tilde{\Phi}{(\cW_{vt})=k}\}}\right.\nnb\\
		&\hspace{25mm}\left.\ind_{\{\SIR_{(0,Y,0)\rightarrow (vt,0,h)}\geq\tau\}}\ind_{\{\tilde{\Phi}(\cW_{vt})=k\}}\right),\label{18}
	\end{align}
where we use the following facts: 1) the rate from the typical IoT to the typical UAV is positive
if and only if it is selected to transmit; 
2) one can write that $ \ind_A=\sum_{k}\ind_{A\cap B_k}\ind_{B_k} $ where $ B_k $ is any partition 
of the sample space; and 3) conditionally on the fact that the number of the IoT device
in the activation window centered at $ vt $, 
$ \cW_{vt}=\left[vt-\frac{w}{2},vt+\frac{w}{2}\right]\times \left[ -\frac{l}{2}, \frac{l}{2}\right] $,
is $k$, the probability for the typical IoT device to be selected to transmit is $ 1/k. $
	\par As a result, combining Eqs. \eqref{17} and \eqref{18}, we obtain the Eq. \eqref{19-1} for the amount of data transmitted from the typical IoT. 
\begin{figure*}
		\begin{align}
	\cD&=\log_2(M)\bE_{\tilde{\Phi}}^0\left[\int_{-\frac{w}{2v}}^{\frac{w}{2v}} \left(\sum_{k=1}^{\infty}\ind_{\{Y \text{ is selected out of $ k $}|\tilde{\Phi}{(\cW_{vt})=k}\}} \ind_{\SIR_{(0,Y,0)\rightarrow (vt,0,h)}\geq\tau}\ind_{\tilde{\Phi}(\cW_{vt})=k}\right)\diff t \right]\nnb\\
	&\ea\log_2(M)\int_{-\frac{w}{2v}}^{\frac{w}{2v}} \bE_{\tilde{\Phi}}^0\left[\sum_{k=1}^{\infty}\ind_{\{Y \text{ is selected out of $ k $}|\tilde{\Phi}{(\cW_{vt})=k}\}} \ind_{\SIR_{(0,Y,0)\rightarrow (vt,0,h)}\geq\tau}\ind_{\tilde{\Phi}(\cW_{vt})=k}\right]\diff t \nnb\\
	&\eb\log_2(M)\int_{-\frac{w}{2v}}^{\frac{w}{2v}}\int_{-\frac{l}{2}}^{\frac{l}{2}}\left(\sum_{k=1}^{\infty}\frac{1}{k} \bE_{\tilde{\Phi}}^0\left[\ind_{\SIR_{(0,y,0)\rightarrow (vt,0,h)}\geq\tau}\ind_{\tilde{\Phi}(\cW_{vt})=k}\right]\right)\frac{\diff y}{l} \diff t\nnb
	\\
	&\ec\log_2(M)\int_{-\frac{w}{2v}}^{\frac{w}{2v}}\int_{-\frac{l}{2}}^{\frac{l}{2}}\left(\sum_{k=1}^{\infty}\frac{1}{k} \bE_{\tilde{\Phi}}^{0}\left[\ind_{\SIR_{(0,y,0)\rightarrow (vt,0,h)}\geq\tau}\right]\bE_{\Phi}^{0}\left[\ind_{\tilde{\Phi}(\cW_{vt})=k}\right]\right)\frac{\diff y}{l} \diff t.\label{19-1}
	\end{align}
		\rule{\textwidth}{0.2pt}	\vspace{-1em}
\end{figure*}
To obtain (a) in Eq. \eqref{19-1}, we use Fubini's theorem. To obtain (b) in Eq. \eqref{19-1}, we use the fact that $ Y $ is an
independent mark of $ \tilde{\Phi} $, where $ Y \sim \text{Uniform}[-\frac{l}{2},\frac{l}{2}]$
and the fact that the probability that $ Y $ is selected out of $ k $ IoT devices is equal to $ 1/k. $
To derive (c) in Eq. \eqref{19-1}, we use the independence of the numbers of points of a Poisson point process in disjoint sets,
where the first term inside the expectation corresponds to the event that the SIR is greater
than $ \tau $ and the second term corresponds to the event that the window of serving UAV has $ k $ points. 
\par Furthermore, the first integrand of Eq. \eqref{19-1} is given by 
	\begin{align}
		&\bE_{\tilde{\Phi}}^{0}(\ind_{\SIR_{(0,y,0)\rightarrow (vt,0,h)}\geq\tau})\nnb\\
		&=\bP_{\tilde{\Phi}}^{0}(\SIR_{(0,y,0)\rightarrow (vt,0,h)}\geq\tau)\nnb\\
		&=\bP_{\tilde{\Phi}}^{0}\left(\frac{pG{\|(-vt,y,-h)\|}^{-{\alpha}}}{\sum_{i\in\bZ}^{\neq 0}pG_i{\|(X_i-vt,Y_i,-h)\|}^{-\alpha}\mathbbm{1}_{\{\tilde{\Phi}(\cW_i)\neq \emptyset\}}}\geq\tau\right)\nnb\\
		&=\bP(G\geq\tau p^{-1} I{(v^2t^2+y^2+h^2)}^{\frac{\alpha}{2}})\nnb\\	&=\left.\sum_{i=0}^{m-1}\frac{(-s)^{i}}{i!}\frac{\diff^i}{\diff s^i}\cL_{I}(s)\right|_{s=\frac{\tau m  (v^2t^2+y^2+h^2)^\frac{\alpha}{2}}{p\Omega}},\label{19-2}
	\end{align}
where $ (X_i,Y_i,0) $ denotes the location of the IoT device selected to transmit in the
window $ \cW_i:=\left[i\mu+vt-\frac{w}{2},i\mu+vt+\frac{w}{2}\right]\times\left[-\frac{l}{2},\frac{l}{2}\right] $
and $ \cL_{I}(s) $ is the Laplace transform of the interference derived in Theorem 1. 
\par Using Slivnyak's theorem, the second integrand of Eq. \eqref{19-1} is given by 
	\begin{align}
		&\bE_{\tilde{\Phi}}^{0}\left[\ind_{\tilde{\Phi}(\cW_{vt})=k}\right]\nnb\\
		&=\bE\left[\ind_{\tilde{\Phi}+\delta_{0} (\cW_{vt})=k}\right]\nnb\\
		&=\bE\left[\ind_{\tilde{\Phi }(\cW_{vt})=k-1}\right]\nnb\\
		&=\bP(\tilde{\Phi} (\cW_{vt})=k-1)=\frac{e^{-\lambda  wl  } {(\lambda  wl)}^{k-1}}{(k-1)!}\label{19-3},
	\end{align}
where we use the fact that intensity of $ \tilde{\Phi} $ is $ \lambda l $.
Eq. \eqref{19-3} holds for $ k=1,2,3,\ldots$ with $ 0!=1. $ 
As a result, by combining Eq. \eqref{19-1},  \eqref{19-2}, and \eqref{19-3}, we have $ \cD $ 
	\begin{align}
		&\log_2(M)\int_{-\frac{w}{2v}}^{\frac{w}{2v}}\int_{-\frac{l}{2}}^{\frac{l}{2}}\left(\underbrace{\sum_{i=0}^{m-1}\frac{(-s)^{i}}{i!}\frac{\diff^i}{\diff s^i}\cL_{I}(s)}_{s=\frac{\tau m  (v^2t^2+y^2+h^2)^\frac{\alpha}{2}}{p\Omega}}\right.\nnb\\
		&\hspace{32mm}\left.\left(\sum_{k=1}^{\infty}\frac{e^{-\lambda  w l  } {(\lambda w l)}^{k-1}}{k(k-1)!} \right)\right)\frac{\diff y}{l} \diff t\nnb\\
		&=\frac{\log_2(M)}{l \cdot 
			\lambda w l}\int_{-\frac{w}{2v}}^{\frac{w}{2v}}\int_{-\frac{l}{2}}^{\frac{l}{2}}\underbrace{\sum_{i=0}^{m-1}\frac{(-s)^{i}}{i!}\frac{\diff^i}{\diff s^i}\cL_{I}(s)}_{s=\frac{\tau m  (v^2t^2+y^2+h^2)^\frac{\alpha}{2}}{p\Omega}}\nnb\\
		&\hspace{32mm}\left(\sum_{k=1}^{\infty}\frac{e^{-\lambda  w l  } {(\lambda  wl)}^{k}}{k!} \right){\diff y} \diff t\nnb\\
		&=\frac{(1-e^{-\lambda wl})\log_2(M)}{\lambda wl^2}\frac{1}{v}\nnb\\
		&\hspace{4mm}\times\int_{-\frac{w}{2}}^{\frac{w}{2}}\int_{-\frac{l}{2}}^{\frac{l}{2}}\underbrace{\sum_{i=0}^{m-1}\frac{(-s)^{i}}{i!}\frac{\diff^i}{\diff s^i}\cL_{I}(s)}_{s=\frac{\tau m  (x^2+y^2+h^2)^\frac{\alpha}{2}}{p\Omega}}\diff y \diff x\nnb.
	\end{align}
This completes the proof. 
\end{IEEEproof}
	\begin{figure*}
	\begin{equation}
	\frac{(1-e^{-\lambda wl})\log_2(M)}{v \lambda wl^2}\!\!\int_{-\frac{w}{2}}^{\frac{w}{2}}\!\int_{-\frac{l}{2}}^{\frac{l}{2}}\!\prod_{k\in\bZ\setminus 0}\!\!\left(e^{-\lambda wl }\!+\!\frac{1-e^{-\lambda w l}}{ w l}\!\!\int_{k\mu-\frac{w}{2}}^{k\mu+\frac{w}{2}}\!\!\int_{-\frac{l}{2}}^{\frac{l}{2}}\frac{\diff v \diff u}{1+\frac{\tau{(x^2+y^2+h^2)}^{\frac{\alpha}{2}}}{{(u^2+v^2+h^2)}^{\frac{\alpha}{2}}}}\right)\diff y \diff x\label{eq:1111}.
	\end{equation}
	
		\rule{\textwidth}{0.2pt}	\vspace{-1em}
\end{figure*}
\begin{example}
With parameters $ m=\Omega=1, $ the mean amount of harvested data is given by Eq. \eqref{eq:1111}.
\end{example}
Here are a few observations on Theorem \ref{T:3}. 1) $ \cD $ is {inversely} proportional to
the speed $ v $; $ \cD\propto v^{-1}. $ If UAVs move faster, each UAV provides a shorter duration 
of active time for each IoT device, and consequently, the average amount of harvested data from
each device decreases. Nevertheless, note that when UAV move faster, each IoT device is
scheduled by a larger number of UAVs per unit time on average. Specifically, the average amount
of harvested data per unit time is given by $ \frac{\cD v}{\mu}  $ since $ v/\mu $ UAVs pass
each IoT device per unit time. 
2)  $ \cD $ is monotonically increasing w.r.t. $  \mu.$ As the distance between UAVs increases,
the interference power decreases, and thus the coverage and rate increase.
Fig. \ref{fig:amountofdatatransmitted} illustrates the evolution of the mean amount of data 
transmitted from the typical IoT device to a UAV in function of $\tau$ and $w/\mu$.
As mentioned in the system model, the size $w$ of the scheduling window is one of the key
control parameters. The impact of this parameter on network performance is
discussed in the next section.
\begin{remark}
	 Both the size of the coverage window and the distance between UAVs could be scaled in order to adapt the density of serving IoT devices.  
\end{remark}

\section{Discussion and Extensions}

\subsection{Mean Rate Formula and Mass Transport}\label{S:5}
In this subsection, we connect the mean data rate of the typical UAV, $ \cR $, given in Section \ref{S:3},
and the mean amount of data transmitted from the typical IoT device, $ \cD $, given in Section \ref{S:4},
by establishing a general relationship between them.

\begin{theorem}\label{T:5}
When the density of IoT devices is high, the mean amount of data harvested from
the typical IoT device per UAV passage, $ \cD $, is linked to the mean rate $\cR$ of the typical UAV
by the relation 
	\begin{align}
	\cD  &\approxeq \frac{\cR}{{\cK}}{} {\cT},\label{MTF}
	\end{align}
where $ {\cK}=\lambda w l $ is the mean number of IoT devices in each window
and $ {\cT}=w/v $ is the mean duration the typical IoT device is in the
activation window of a given UAV.
\end{theorem}

\begin{IEEEproof}
When the density of IoT device is high enough, $ 1-e^{-\lambda wl} \approxeq 1 $.
Then, using Eqs.\eqref{eq:r} and \eqref{eq:T3}, we can write 
	\begin{equation}
	{\cD}\approxeq{\frac{1}{\lambda w l}}{\frac{w}{v}}{\cR}= \frac{\cR}{{\cK}}{\cT}.
	\end{equation}
This completes the proof. 
\end{IEEEproof}
\begin{remark}
In Appendix \ref{A:1}, the above mean formula is proved again 
based on the mass transport principle \cite{baccelli2010stochasticvol2}
on a unimodular weighted graph. In the graph, UAVs and the activated IoT devices are 
vertices of the graph and links from the IoT devices to UAVs are the edges of the graph.
This shows that the above relationship holds mainly due to the following two key reasons: 
1) the IoT and UAV point processes are jointly stationary and 2) only one IoT device per window
is scheduled to transmit and IoT devices are scheduled for the same amount of time, on average.
Consequently, the same linear relationship will hold for other and more general network scenarios
(e.g., other than Poisson but stationary IoT device point processes,
other than Nakagami fading models, etc.). 
\end{remark}

\subsection{Performance Optimization w.r.t. Activation Window}\label{S:3A}

\begin{figure}
	\centering
	\includegraphics[width=1\linewidth]{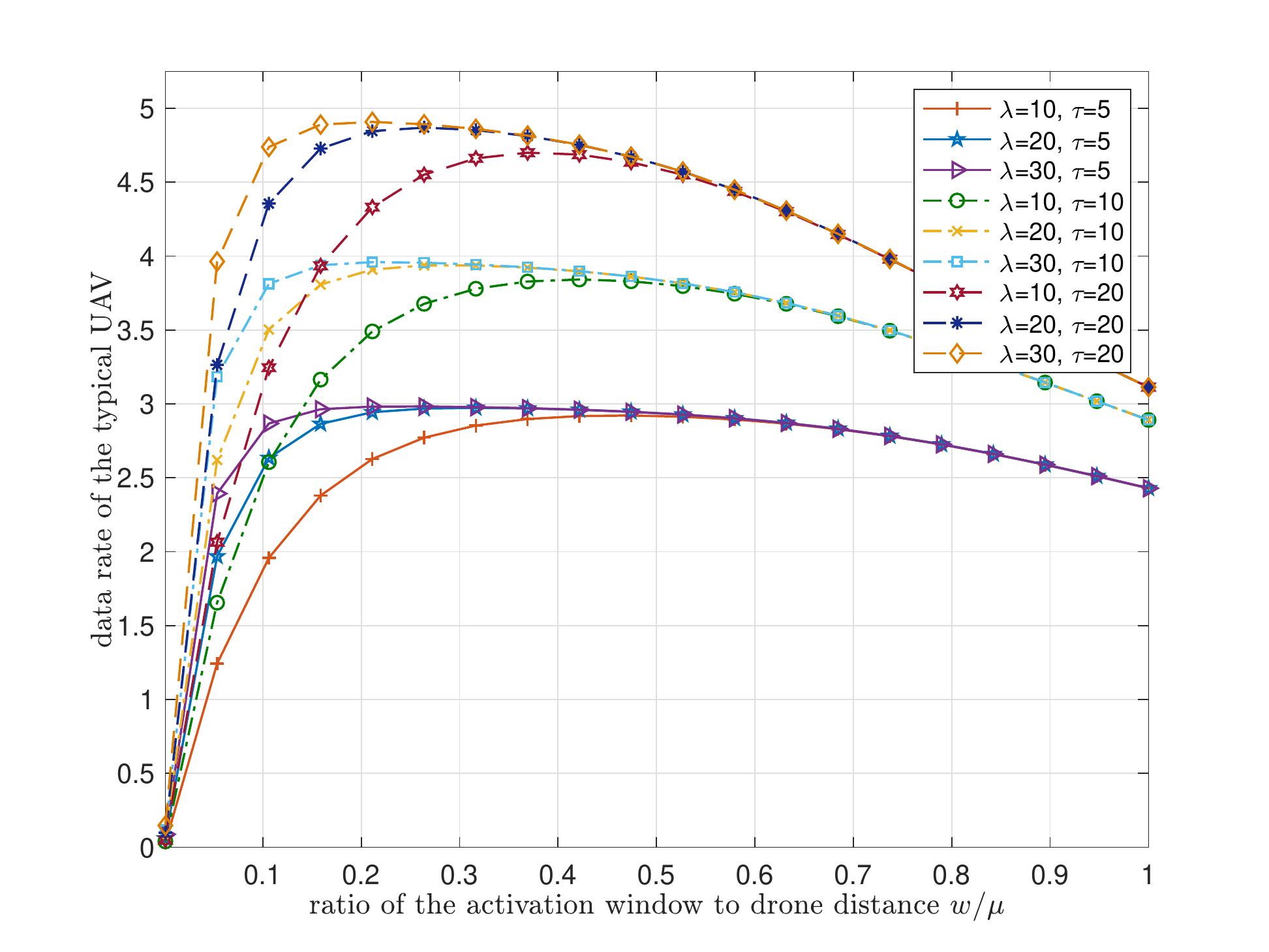}
	\caption{Illustration of the data rates at the typical UAV.  The maximum size of activation window is equal to $ \mu=2 \text{ km} $ We consider $ \alpha=4 $, $ h=0.2 $ km, Nakagami fading parameters $ m=\Omega=1 $, and the modulation rate $ M=2^{\floor{\log_2(1+\tau)}}. $}	
	\label{fig:optimizationwindow}
\end{figure}


The activation windows determine the area where IoT devices are scheduled for uplink transmissions.
In this paper, these windows are modeled by a collection of rectangles on the plane.
The length of each window $ w, $ is a control parameter that should be configured to
achieve the best network performance. We recall that we assume that the distance between UAVs 
is larger than the length of the activation ($ \mu>w $).  
We show below that $w$ is one of the key optimization parameters of the proposed
motion-based harvesting architecture.

Let us discuss the impact of the activation windows on the network performance.
If $ w $ is too big, i.e., $ w\approxeq \mu $, the key advantage of the proposed architecture
vanishes as the idea is to provide universal coverage with a number of UAVs smaller
than the number of static base stations that would cover the whole plane.
In addition, a larger window leads to a larger number of IoT devices scheduled for the uplink 
transmissions in the network. Consequently, the average distance from IoT devices to their
corresponding UAVs also increases, as well as the interference at any UAV. 
On the other hand, if the window $ w $ is too small, it is very likely that
activation windows are empty of IoT devices. Therefore, the upload data rate
becomes negligible. Fig. \ref{fig:optimizationwindow} illustrates the dependency of
the mean data rate of the typical UAV $ \cR $ w.r.t. $w$, normalized by the inter-UAV distance.
In this figure, we use the parameters $ \mu=2 \text{ km} , h=0.25 \text { km}$ and $ \alpha=4.
$ When $ w=0 $ km, we have $ \cR\approxeq 0. $ For smaller values of the window size,
the data rate increases as the window size increases. For larger values, the data rate decreases
as the window size increases. There is a unique value of $w$ where the data rate achieves its maximum.
For densities $ \lambda= 10,20,$ and $30 /\text{km}^2 $, the corresponding optimum window sizes
are roughly $ 0.2\mu $, $ 0.3\mu $, and $ 0.4 \mu. $ The figure clearly demonstrates
that the optimal size of activation window $ w^\star $ depends on the density $ \lambda. $
Generally speaking, for a higher density of IoT devices, a smaller activation window is
required to optimize the data rate. The optimal size of window can be numerically found by Eq. \eqref{eq:opti}
\begin{figure*}
	\begin{align}
	\argmax_{0\leq w\leq \mu}\frac{(1-e^{-\lambda w l})}{ w l}\int_{-\frac{w}{2}}^{\frac{w}{2}}\!\int_{-\frac{l}{2}}^{\frac{l}{2}}\prod_{k}^{\bZ\setminus 0}\left(\!e^{-\lambda wl}+\frac{1-e^{-\lambda w l}}{w l}\int_{k\mu-\frac{w}{2}}^{k\mu+\frac{w}{2}}\!\int_{-\frac{l}{2}}^{\frac{l}{2}}\!\frac{\diff v \diff u}{\left(1+\frac{\Omega\tau{(x^2+y^2+h^2)}^{\frac{\alpha}{2}}}{m{(u^2+v^2+h^2)}^{\frac{\alpha}{2}}}\right)^m}\!\right)\!\diff y\diff x.\label{eq:opti}
	\end{align}
		\begin{equation}\label{Laplace2D}
	e^{-sN_0}\prod_{i \in\bZ}\left(e^{-\lambda w l}+\frac{1-e^{-\lambda w l }}{wl}\int_{i\mu-\frac{w}{2}}^{i\mu+\frac{w}{2}}\int_{-\frac{l}{2}}^{\frac{l}{2}}\left(\frac{1}{1+\frac{sp\Omega m^{-1}}{{(x^2+y^2+h^2)}^{{\alpha}/{2}}}}\right)^m\diff y \diff x\right).
	\end{equation}
	\rule{\textwidth}{0.2pt}	\vspace{-1em}
\end{figure*}
where the constant $ \log_2(1+M) $ is removed from $ \cR $  for the moment
because it cannot change the optimal window size.
Solving the above optimization exactly, e.g., KKT condition \cite{boyd2004convex},
is beyond of the scope of this paper.

\subsection{Interference-limited Architecture}

\begin{figure}
	\centering
	\includegraphics[width=1\linewidth]{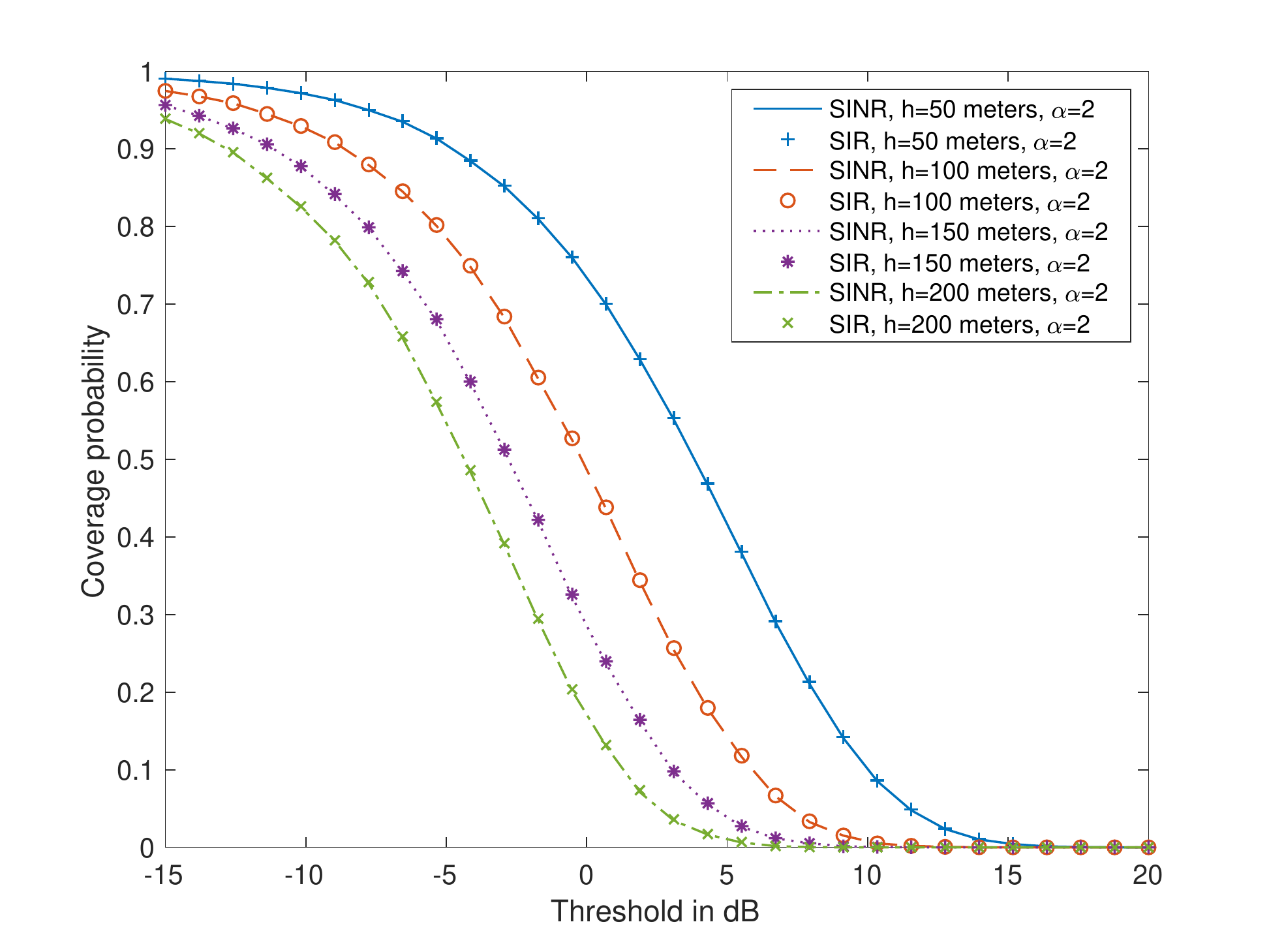}
	\caption{The simulated probability that the typical SINR and SNR is greater than some threshold $ \tau. $ I consider $ p=23\text{ dBm}, \alpha=2, w,l=100\text{ meters}, \mu=200 \text{meters}, \sigma^2=-104 \text{ dBm}$. }
	\label{fig:sinrvssir}
\end{figure}

\begin{figure}
	\centering
	\includegraphics[width=1\linewidth]{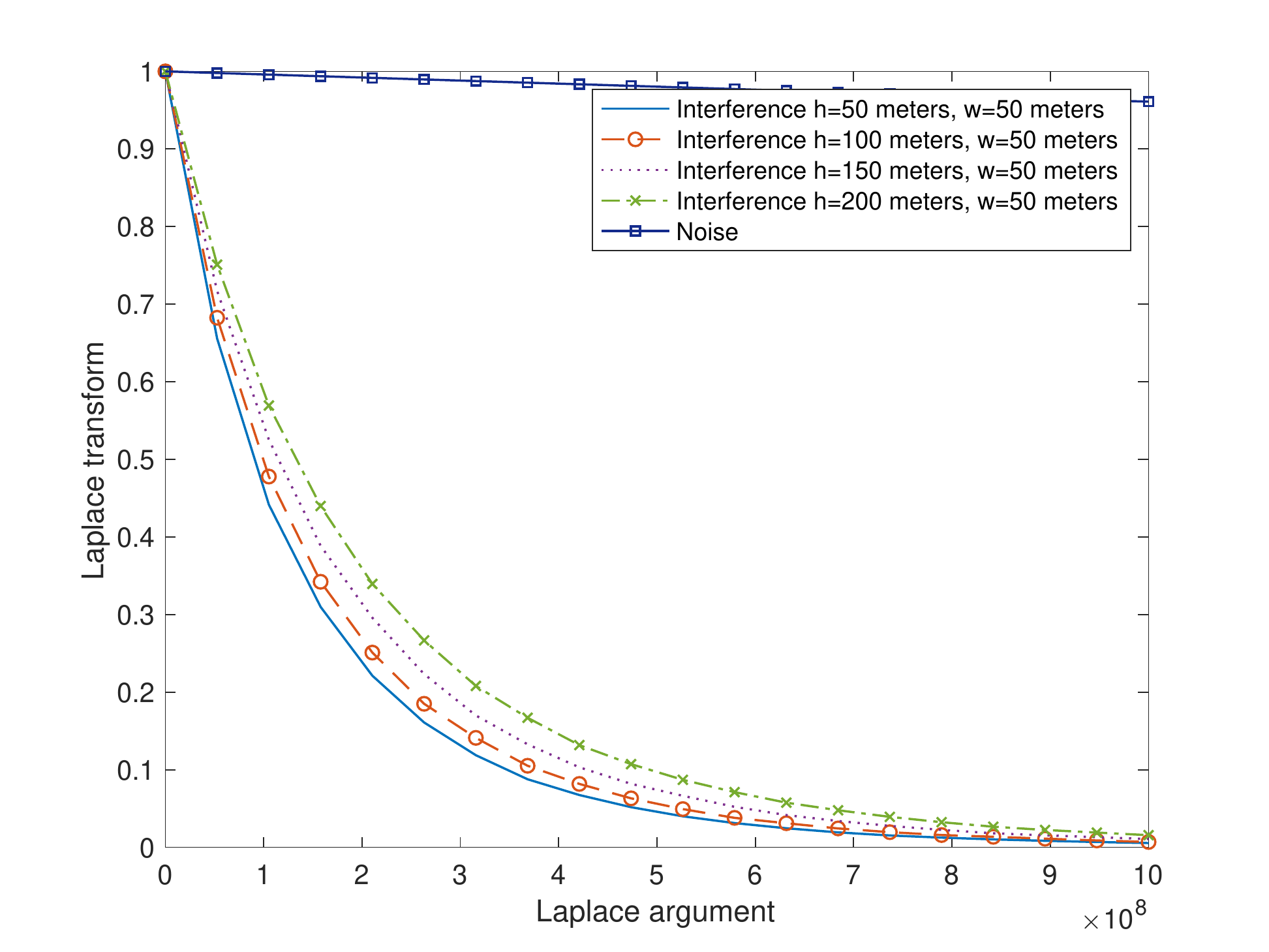}
	\caption{Illustration of Laplace transform of the interference and noise. We consider $ p=23 $ dBm, $ w,l=100 $ meters, $ \mu=200 $ meters, $ \sigma^2=-104 $ dBm.}
	\label{fig:laplacetransform}
\end{figure}

In the coverage analysis, the thermal noise at UAVs is ignored. The major difference the proposed network with the conventional UAV networks is that the proposed data harvesting architecture is based on a fleet of UAVs in a single line (or in parallel lines for the 2-D case that will be provided shortly). Thus, the activation windows of UAVs create a non-negligible amount of interference at the typical UAV and the proposed network is interference-limited.  Table \ref{Ta:3} describes the network parameters used for system-level simulations of the proposed network. In Fig. \ref{fig:sinrvssir}, the SINR and SIR coverage probabilities are illustrated. Each SINR and corresponding SIR are almost identical, which indicates that the noise power at the typical UAV is very small compared to the interference power at the typical UAV for the given parameters. In Fig. \ref{fig:laplacetransform}, the Laplace transforms of the interference and noise power are illustrated. The Laplace transform of the interference plus thermal noise is given by Eq. \eqref{Laplace2D} where $ N_0 $ denotes thermal noise. Note the Laplace transform of the interference is less than the Laplace transform of the noise power, which implies that the interference power is greater than the noise power for the given parameters.

\begin{table}
	\caption{Simulation parameters}\label{Ta:3}
	\centering
	\begin{tabular}{|c|c|}
		\hline 
		Bandwidth	& 10 MHz  \\ 
		\hline 
		Density of IoT devices	& $ 10^5 $/km$ ^2 $ \\ 
		\hline 
		Transmit power	&  23 dBm \\ 
		\hline 
		Noise power density		&  -174 dBm/Hz\\ 
		\hline 
		UAV Altitude & $ \{50,100,150,200\} $ meters \\ 
		\hline
		Carrier frequency & $ 1 $ GHz \\
		\hline
		Path loss exponent & $ 2 $\\ 
		\hline
		Window dimension & $ \{50 \times 100 \text{ m}^2, 100 \times 100 \text{ m}^2 \}$ \\
		\hline
		Inter-UAV distance & $ 200 \text{ meters}$ \\
		\hline
	\end{tabular} 
\end{table}

\subsection{Two-Dimensional Extensions}\label{S:twodim}
\begin{figure}
	\centering
	\includegraphics[width=1\linewidth]{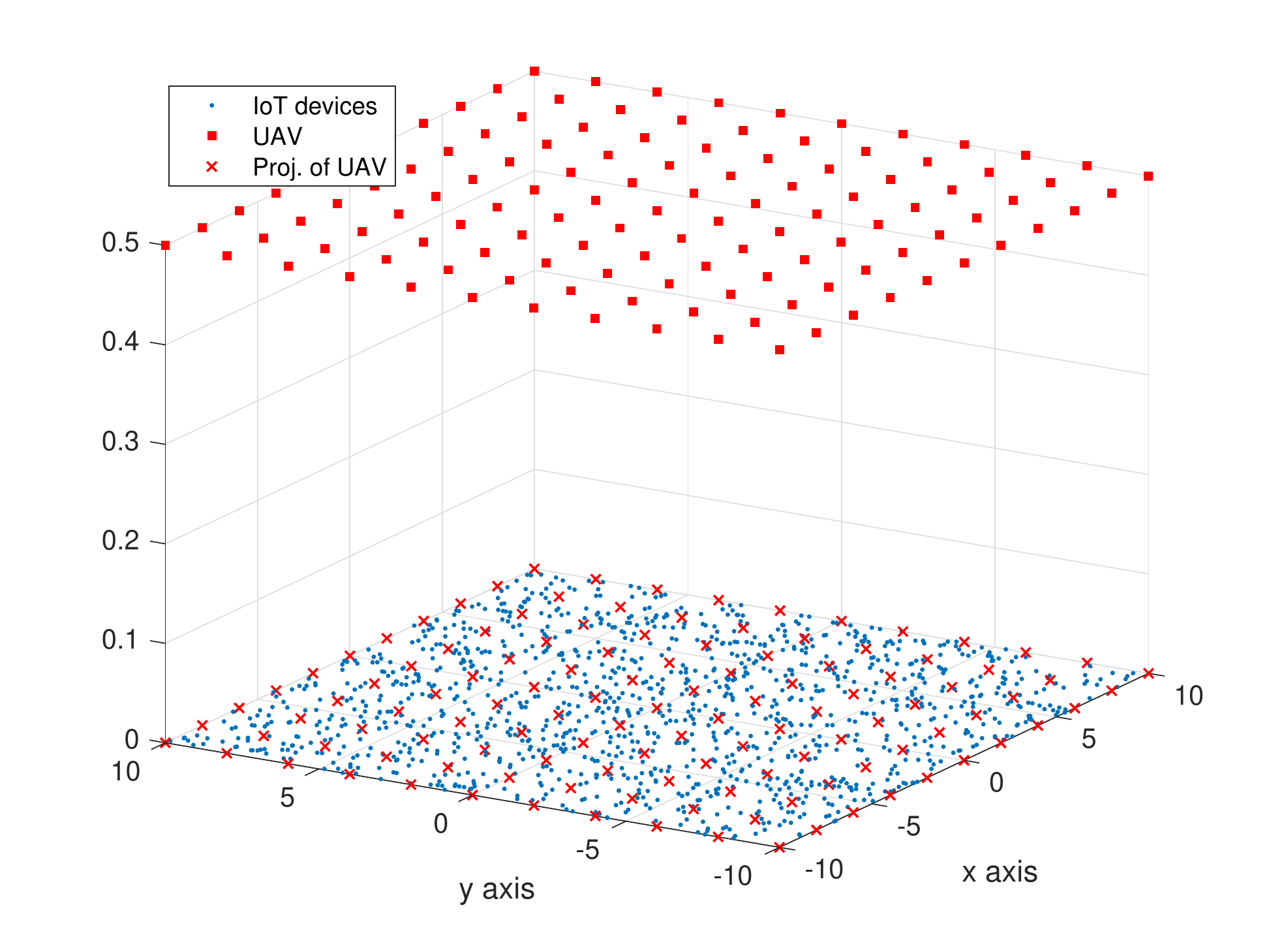}
	\caption{2-D extension of the proposed linear model.}
	\label{fig:droneiot3d}
\end{figure}

The linear model discussed so far was motivated by applications
where UAVs hover or travel, e.g., over a street in a coordinated way
to collect data from roadside units. 
As discussed in the motivation section, for other applications, 
there is a need for 2-D UAV networks covering 
in an intermittent but universal way the whole Euclidean plane.
In the 2-D extension discussed in this subsection,
we assume that IoT devices are distributed according to a {planar} Poisson point process
$ \hat{\Phi} $ with spatial density $ \hat{\lambda}$.
There is a collection of parallel linear trajectories with a periodically spaced fleet of UAVs on each.
Each UAV fleets moves along its trajectory with a fixed speed $ v. $ Hence
\begin{align}\label{eq:2-dmodel}
\hat{\Psi}(t)=\sum_{i,j\in\bZ^2}\delta_{(i\mu,j\nu,h)+(U,V,0)+(vt,0,0)},
\end{align}
where $ U \sim \text{Uniform}[-\mu/2,\mu/2]$, $ V\sim\text{Uniform}[-{\nu}/{2},{\nu}/{2}] $,
and $ \nu $ is the distance between adjacent UAV trajectories. Similar to the strip UAV model,
the randomized shift modeled by $ U,V $ ensures the stationarity of the 2-D UAV process
$ \hat{\Psi}(t) $. The activation window is modeled by $ w  $ by $ l. $
Note that if the proposed architecture is meant to provide universal coverage,
we will have $ w\leq \mu $ and $ l = \nu. $
Fig. \ref{fig:droneiot3d} illustrates the 2-D model with parameters: 
$ \mu=\nu=l= 2 \text{ km}$. 
Note that other (an possibly better) variants can be considered like, e.g.,
having an hexagonal grid of UAVs rather than the square one considered in the illustration.

The methodology developed for the 1-D case can be extended step by step.
\begin{figure*}
		\begin{equation}
	\cL_{{N_2}}(s)=\prod_{(i,j)\in\bZ^2\setminus \{(0,0)\}}\left(e^{-\hat{\lambda} w l}+\frac{1-e^{-\hat{\lambda} w l }}{wl}\int_{i\mu-\frac{w}{2}}^{i\mu+\frac{w}{2}}\int_{jl-\frac{l}{2}}^{jl+\frac{l}{2}}\frac{1}{\left(1+\frac{sp\Omega m^{-1}}{{(x^2+y^2+h^2)}^{\frac{\alpha}{2}}}\right)^m}\diff y \diff x\right).\label{eq:dddd}
	\end{equation}
			\begin{equation}
	\cL_{I_2}(s)=\prod_{(i,j)\in\bZ^2}^{\neq (0,0)}\left(e^{-\hat{\lambda} w l}+\frac{1-e^{-\hat{\lambda} w l}}{ w l}\int_{i\mu-\frac{w}{2}}^{i\mu+\frac{w}{2}}\int_{jl-\frac{l}{2}}^{jl+\frac{l}{2}}\left(\frac{1}{1+\frac{sp\Omega }{m{(x^2+y^2+h^2)}^{-\frac{\alpha}{2}}}}\right)^{m}\diff y\diff x\right).\label{26}
	\end{equation}
		\rule{\textwidth}{0.2pt}	\vspace{-1em}
\end{figure*}
\begin{corollary}\label{P:1}
	The Laplace transform of the uplink shot-noise process of the typical UAV in the 2-D model is given by  Eq. \eqref{eq:dddd}.  Furthermore, the shot-noise process at the typical UAV is time-invariant.
\end{corollary}
\begin{IEEEproof}
Under the Palm distribution of the UAV point process $ \hat{\Psi} $,
the typical UAV exists at $  (0,0,h).$  The 2-D activation window $ \hat{W} $ is given by 
	\begin{align*}
	\hat{\cW}&=\hat{\cW}_{i,j}\\
	&=\bigcup_{(i,j)\in\bZ^2} \left[i \mu  -\frac{w}{2},i \mu  +\frac{w}{2}\right]\times \left[j \nu  -\frac{l}{2}, j \nu +\frac{l}{2}\right],
	\end{align*}
	where $ \mu $ is the distance between UAVs on the same trajectory and $ \nu $ indicates the distance between UAV trajectories; $ w\leq\mu  $ and $ l \leq \nu $. The shot-noise process at the typical UAV is 
	\begin{align*}
	{{N_2}}=\sum_{(X_{i,j},Y_{i,j})}\!\!\!\! p G \|(X_{i,j},Y_{i,j},0)-(0,0,h)\|^{-\alpha}\mathbbm{1}_{\{\hat{\Phi}(\hat{\cW}_{i,j})\neq \emptyset\}},
	\end{align*}
where $ (X_{i,j},Y_{i,j}) $ are the $ x,y $ coordinates of the access-granted IoT devices,
if any, in window $ \hat{\cW}_{i,j} $. Due to the Poisson property, the locations of
the access-granted IoT device is uniformly distributed in its corresponding window. 
Similar to the proof of Theorem \ref{T:1}, we can write $ \cL_{{N}_2}(s) $ as follows:
	\begin{align}
	&\bE_{\hat{\Psi}}^0\left[\prod_{(X_{i,j},Y_{i,j})}^{\in\hat{\cW}_{i,j}}\left(\bP(\hat{\Phi}(\hat{\cW}_{i,j})=\emptyset)\right.\right.\nnb\\
	&\hspace{8mm}+\left.\left.\bP(\hat{\Phi}(\hat{\cW}_{i,j})\neq \emptyset)\!\bE\!\left[\!e^{-s p G{\|(X_{i,j},Y_{i,j},0)-(0,0,h)\|}^{-\alpha}}\right] \right)\right]\nnb\\
	&=\bE_{\hat{\Psi}}^0\left[\prod_{(X_{i,j},Y_{i,j})}^{\in\hat{\cW}_{i,j}}\left(e^{-\hat{\lambda} w l}\right.\right.\nnb\\
	&\hspace{10mm}\left.\left.+\left({1-e^{-\hat{\lambda} w l}}\right)\int_{\text{supp}(G)}\hspace{-8mm}e^{-spg{(Y_{i,j}^2+X_{i,j}^2+h^2)}^{-\frac{\alpha}{2}}}f_G(g)\diff g\right)\right]\nnb\\
	&\ea\prod_{(i,j)\in\bZ^2}\left(e^{-\hat{\lambda} w l}\right.\nnb\\
&\left.+\frac{1-e^{-\hat{\lambda} w l}}{ w l}\!\!\int_{i\mu-\frac{w}{2}}^{i\mu+\frac{w}{2}}\!\!\!\int_{jl-\frac{l}{2}}^{jl+\frac{l}{2}}\!\left(\frac{1}{1+\frac{sp\Omega }{m{(x^2+y^2+h^2)}^{-\frac{\alpha}{2}}}}\right)^{m}\!\!\!\!\diff y\diff x\!\!\right)\nnb,
	\end{align}
where we obtain (a) from the fact that the access-granted IoT devices are
independent and they are uniformly distributed in each window.  
\end{IEEEproof}

\begin{example}
The Laplace transform of the interference of the typical UAV is given by Eq. \eqref{26}.
\end{example}
\par The Laplace transform of the 2-D shot-noise process can be used to compute
the coverage probability and the data rate of the typical UAV. 

\begin{corollary}
The coverage probability of the typical UAV is given by 
	\begin{align*}
	\frac{1-e^{-\hat{\lambda} w l}}{ w l}\int_{-\frac{w}{2}}^{\frac{w}{2}}\int_{-\frac{l}{2}}^{\frac{l}{2}}\underbrace{\sum_{i=0}^{m-1}\frac{(-s)^{i}}{i!}\frac{\diff^i}{\diff s^i}\cL_{I_2}(s)}_{s\leftarrow\frac{\tau m  (x^2+y^2+h^2)^\frac{\alpha}{2}}{p \Omega}}\diff y\diff x.
	\end{align*}
	Similarly, the data rate at the typical UAV is given by 
	\begin{align*}
	\frac{\log_2(M)(1-e^{-\hat{\lambda} w l})}{ w l}\int_{-\frac{w}{2}}^{\frac{w}{2}}&\!\int_{-\frac{l}{2}}^{\frac{l}{2}}\underbrace{\sum_{i=0}^{m-1}\frac{(-s)^{i}}{i!}\frac{\diff^i}{\diff s^i}\cL_{I_2}(s)}_{s=\frac{\tau m  (x^2+y^2+h^2)^\frac{\alpha}{2}}{p \Omega}}\diff y\diff x.
	\end{align*}
In both Eqs, $ \cL_{I_2}(s) $ denotes the Laplace transform of the interference provided in Eq. \eqref{26}.
\end{corollary}
\begin{IEEEproof} The proof is similar to the proof of Theorem 2.
\end{IEEEproof}

\section{Conclusion}\label{S:7}
This paper analyzes the performance of data harvesting architectures based on static IoT
devices and a fleet of UAVs with coordinated motion. In the proposed framework, UAVs harvest
delay-tolerant data from surface IoT devices, inside their activation windows.
Due to the fleet's inherent mobility, the proposed architecture can provide universal coverage
for the widely distributed IoT device even when the activation window size is smaller
than the instantaneous coverage area. By leveraging the joint stationarity of the spatial model,
we derived the coverage probability and the mean data rate of the typical UAV.
Similarly, in order to assess the harvesting capability
of the proposed architecture, we derived a formula for the mean amount of data uploaded 
from the typical IoT device to a UAV. This metric was then linked to the data rate of the typical UAV,
thanks to the mass transport principle. 
The key features of the proposed architecture can be summarized as follows:
\begin{itemize}
	\item Its instantaneous coverage area (activation windows) can be way smaller than 
the total coverage area at all times, which allows for cost reductions (to the expense of delay).
	\item It has the potential of providing large-scale universal connectivity with a
bounded connectivity delay, both in 1 and 2-D.
	\item Its key performance metrics are known in closed or integral form.
        \item Its harvesting capacity can be maximized by selecting some optimal activation window.
\end{itemize}

\appendices
\section{Proof of Theorem \ref{T:5}}\label{A:1}
We prove Theorem \ref{T:5} using the mass transport principle on a stationary graph, defined on jointly stationary point processes. Let $ \Phi_1$ and $ \Phi_2 $ be arbitrary stationary point processes with intensity $ \lambda_1 $ and $ \lambda_2, $ respectively. Let $ G $ denote a weighted directed graph with weights equal to the amount of mass transported from the points of $ \Phi_1 $ to the points of $ \Phi_2 $. Assuming that the graph and the weights are factors of $ (\Phi_1,\Phi_2), $ we have the following mass transport formula:
\begin{equation}\label{MTP}
\frac{\lambda_1}{\lambda_1+\lambda_2}\bE_{\Phi_1}^0[G^+(0)]=	\frac{\lambda_2}{\lambda_1+\lambda_2}\bE_{\Phi_2}^0[G^-(0)],
\end{equation}
where $ G^+(0) $ and $ G^-(0) $ denote the mass out of the origin under $ \bP_{\Phi_1}^0 $, and the mass toward the origin under $ \bP_{\Phi_2}^0 $, respectively.

\par Consider the graph whose vertices are given by $ (\Psi,\Phi') $; the UAV point process $ \Psi $ and the corresponding IoT point process $ \Phi' $ inside the windows of $ \Psi $. Note that $ \Phi'\subset \Phi $ and $ (\Psi,\Phi') $ are joint stationary. Due to the Poisson property, the density of $ \Psi $ is  $ \frac{1}{\mu} $ and the density of $ \Phi' $ is equal to $ \frac{\lambda w l}{\mu}$.  Under the TDMA scheduling of UAVs, the weights of edges are the instantaneous data rates from the points of $ \Phi' $ to the points of $ \Psi' $. By the mass transport in Eq. \eqref{MTP}, we have
\begin{equation}
\frac{\frac{\lambda w l}{\mu}}{\frac{1}{\mu}+\frac{\lambda w l}{\mu}}\cR_{out}=\frac{\frac{1}{\mu}}{\frac{1}{\mu}+\frac{\lambda w l}{\mu}}\cR_{in},\nnb
\end{equation}
where $ \cR_{out} $ is the instantaneous data rate from the points of $ \Phi' $ and $\cR_{in}$ is the instantaneous data rate at the points of $ \Psi $, namely, the instantaneous data rate seen from the typical UAV. Based on the mass transport principle, we have 
\begin{align}
\lambda wl \cR_{out}&=\cR_{in}=\cR.\label{266}
\end{align} 
\par On other hand, due to the fact that IoT devices are covered by UAVs repeatedly, the amount of data $ \cD $ transmitted from the typical IoT device per UAV passage is given by 
\begin{align}
\cD&={\cR_{out}}\uta{\frac{w}{v}}+0 \cdot \utb{\frac{\mu-w}{v}}\nnb\\
&=\cR_{out}\frac{w}{v}\ec\cR_{in} \frac{1}{\lambda w l}\frac{w}{v}=\cR\frac{\cT}{\cK},\nnb
\end{align} 
where (a) is the amount of time that the each IoT device is inside each window and (b) is the amount of time that each IoT device is outside of each window. To derive (c), we use Eq. \eqref{266}. This completes the proof. 

\bibliographystyle{IEEEtran}

\bibliography{MCN}

\end{document}